\documentclass[twocolumn]{aastex701}

\usepackage[version=4]{mhchem}
\usepackage{lipsum}
\usepackage{pifont}%
\usepackage{makecell}
\usepackage{tabularx}
\usepackage{physics}
\let\tablenum\relax 
\usepackage{siunitx}
\usepackage{amsmath}

\usepackage{graphicx}

\begin{document}
\title{Three outstanding physical questions for K2-18 b and other temperate sub-Neptunes}

\author[0000-0002-8163-4608]{Shang-Min Tsai}
\affiliation{Institute of Astronomy \& Astrophysics, Academia Sinica, Taipei 10617, Taiwan}
\email[show]{smtsai@asiaa.sinica.edu.tw} 

\author[0000-0001-6096-7772]{Piero Ferrari}
\affiliation{HFML-FELIX, Nijmegen 6525 ED, The Netherlands.} 
\email{piero.ferrariramirez@ru.nl} 

\author[0009-0003-1024-8931]{Mats Kuipers}
\affiliation{HFML-FELIX, Nijmegen 6525 ED, The Netherlands.}
\email{mats.kuipers@ru.nl}

\author[0000-0002-0746-1980]{Jacob Lustig-Yaeger}
\affiliation{JHU Applied Physics Laboratory, 11100 Johns Hopkins Rd, Laurel, MD 20723, USA}
\email{Jacob.Lustig-Yaeger@jhuapl.edu} 

\author[0000-0003-2944-0600]{Arnav Agrawal}
\affiliation{JHU Applied Physics Laboratory, 11100 Johns Hopkins Rd, Laurel, MD 20723, USA}
\email{Arnav.Agrawal@jhuapl.edu} 

\author[0000-0002-2828-0396]{Sean Jordan}
\affiliation{ETH Zurich, Institute for Particle Physics \& Astrophysics, Wolfgang-Pauli-Str. 27, 8093 Zurich, Switzerland}
\email{jordans@ethz.ch} 

\author[0000-0001-6842-9004]{Bart Oostenrijk}
\affiliation{Deutsches Elektronen-Synchrotron DESY, Hamburg, Germany}
\email{Bart.oostenrijk@desy.de} 

\author[0000-0002-4274-365X]{Laura Pille}
\affiliation{Deutsches Elektronen-Synchrotron DESY, Hamburg, Germany}
\email{laura.pille@desy.de} 

\author[0000-0002-8199-9818]{Sophia R. Vaughan}
\affiliation{Max-Planck-Institut f\"ur Astronomie, K\"onigstuhl 17, 69117 Heidelberg, Germany}
\email{sovaughan@mpia.de}

\author[0000-0002-2949-2163]{Edward W. Schwieterman}  
\affiliation{Department of Earth and Planetary Sciences, University of California, Riverside, CA, USA}
\affiliation{Virtual Planetary Laboratory, University of Washington, Seattle, WA 98195}
\affiliation{Blue Marble Space Institute of Science, Seattle, WA, USA}
\email{eschwiet@ucr.edu} 

\author[0000-0001-8407-4020]{Laurens B. F. M. Waters} 
\affiliation{Department of Astrophysics, IMAPP, Radboud University, Nijmegen, The Netherlands}
\affiliation{SRON Leiden, Niels Bohrweg 4, 2333 CA Leiden, The Netherlands}
\email{rens.waters@ru.nl} 


\begin{abstract}
Recent transmission spectra of the temperate sub-Neptune K2-18 b obtained with JWST have attracted significant attention. Debates have quickly arisen over the interpretation of the spectral data, particularly the recent MIRI observation where dimethyl sulfide (DMS) and dimethyl disulfide (DMDS) are claimed.  Here we revisit K2-18 b as a case study to examine several key questions that are also broadly relevant to the temperate sub-Neptune population: i) Can the low water abundance be reconciled with water clouds driven by orbital eccentricity? ii) Are the observed and non-observed atmospheric compositions mutually consistent? iii) Is it kinetically possible to produce DMS under sub-Neptune conditions? To address these questions, we couple climate and photochemical models to obtain self-consistent climate-photochemistry states for K2-18 b with a moderate orbital eccentricity of 0.2, as suggested by radial-velocity (RV) measurements. In addition, we present new laboratory measurements of DMS and DMDS infrared opacities by HFML-FELIX and compile updated \ce{C2H6} (ethane) opacities that include weak overtone bands. Our results support a sub-Neptune interpretation without invoking DMS,  and a high-metallicity envelope without enhanced deep water abundance.
\end{abstract} 


\keywords{\uat{Exoplanet atmospheres}{487} --- \uat{Exoplanet atmospheric composition}{2021} --- \uat{Transmission spectroscopy}{2133}}

\section{Introduction}

The Kepler Space Telescope campaigns reveal that sub-Neptunes and super-Earths around M dwarfs represent the most common planet populations around the most common type of stars \citep{Hsu2019,HardegreeUllman2025}. Their interior compositions, inferred from intermediate bulk densities ($\rho_{\mathrm{H_2}} < \rho < \rho_{\oplus}$) derived from mass and radius measurements, remain highly degenerate. One promising avenue for breaking this degeneracy is atmospheric characterization, as several principal gases depend on thermochemical recycling on sub-Neptunes or the interaction with the planetary surface or interiors \citep{Yu2021,Hu2021,Tsai2021b,Shorttle2024,Luu2024,Huang2024,Nixon2025}. 

Currently, two theories have been proposed to explain the radius distribution of sub-Neptunes \citep{Bean2021}. One interpretation,  following Occam's razor, assumes that planets form with the same rocky core, where sub-Neptunes represent the population where the size of the envelope becomes comparable to the core and stable against atmospheric loss \citep{Owen2013}. The term ``gas-dwarfs" is used to describe the sub-Neptune population formed in such a way.

The other explanation is that a large fraction of sub-Neptunes have water-rich interiors, following the early theoretical studies \citep{Kuchner2003} and informed by the density gap analysis \citep{Zeng2019,Luque2022}. One further hypothesis is that some of these temperate, water-rich sub-Neptunes can host water oceans underlying an \ce{H2}-atmosphere, referred to as ``Hycean" (Hydrogen-ocean) planets \citep{Madhusudhan2021}. 

Current JWST observations target several temperate-to-warm sub-Neptunes with equilibrium temperatures ($T_{\mathrm{eq}}$) spanning roughly 200 to a few hundred K. Their \ce{H2}-dominated atmospheres are favorable for characterization through transit spectroscopy and reveal a diverse range of compositions \citep{Madhusudhan2025b}. Discovered in the K2 campaign, K2-18 b is an archetype of temperate sub-Neptunes. To zeroth order, it lies within the habitable zone, with an equilibrium temperature of approximately $280 \pm 19$ K, taking into account observational uncertainties and without assuming any Bond albedo \citep{Benneke2019,Sarkis2018,Howard2025}.

Previous transmission observations of K2-18 b with the Hubble Space Telescope (HST) attributed the 1–1.7 $\mu$m near-infrared features to \ce{H2O} \citep{Benneke2019,Tsiaras2019}. The absorption degeneracy within the narrow HST/WFC3 bandpass between \ce{CH4} and \ce{H2O} was later pointed out by \citet{Blain2021}, with the weak lines of \ce{CH4} suggested to more likely dominate the observed absorption \citep{Bezard2022}. With JWST, the broader wavelength coverage helps break this degeneracy and indicates that the feature is indeed primarily attributed to \ce{CH4}. The first JWST transmission spectra of K2-18 b obtained with NIRISS and NIRSpec reported detections of \ce{CO2} and \ce{CH4}, with non-detections of \ce{H2O} and \ce{NH3} \citep{Madhusudhan2023}.

Since \ce{NH3} is expected to be robustly maintained on a sub-Neptune by rapid thermochemistry in the deep atmospheres \citep{Yu2021,Hu2021,Tsai2021b}, \cite{Madhusudhan2023} interpreted K2-18 b as a Hycean planet where \ce{NH3} is dissolved into the water ocean. Currently, the debate over the Hycean versus sub-Neptune interpretation is not fully resolved. From the climate perspective, the greenhouse effect of hydrogen makes the Hycean scenario challenging, as it is difficult to prevent the water ocean from entering a supercritical state without reflective aerosols \citep{Scheucher2020,Pierrehumbert2023,Innes2023,Leconte2024}. From the photochemical perspective, \ce{CH4} is expected to be depleted by photodissociation on a timescale shorter than the system's age on a Hycean world with a thin atmosphere \citep{Wogan2024}. The \ce{CH4} depletion can be suppressed with a few tens of bars of surface pressure under a weak UV environment (at least 75$\%$ of a low‑activity GJ 436-like UV flux; \citealt{Cooke2024}). Alternatively, a gas-rich sub-Neptune with solar metallicity naturally produces \ce{CH4} and \ce{CO2} without fine-tuning. However, the lack of \ce{NH3} requires a reduced magma ocean beneath the envelope that dissolves \ce{NH3}, which might be challenging to confirm observationally \citep{Lichtenberg2025}. From the planet formation perspective, it remains unclear whether an \ce{H2O}-rich interior can produce the intermediate bulk density, as most of the \ce{H2O} partitioned into the interior likely does not remain in an isolated \ce{H2O} layer \citep{Dorn2021,Vazan2022}.

\cite{Madhusudhan2023} additionally reported tentative detection of dimethyl sulfide (DMS), which is a metabolic product of many marine phytoplankton and algae on Earth. Its overwhelmingly biogenic source and short photochemical timescale make it a strong biosignature gas \citep{Shawn2011,Schwieterman2018}, particularly more observable for anoxic biospheres. \cite{Tsai2024b} demonstrated that a high level of biological activity (about 20$\times$ modern Earth's value) is plausible to maintain detectable DMS in a hypothetical Hycean planet, as UV shielding by organic sulfur gases suppresses DMS photolysis, yet its NIRSpec detection might be highly degenerate with that of methane due to the dominance of \ce{CH4} in the 3–4 $\mu$m region \cite[e.g., ][]{Niraula2025}. \cite{Madhusudhan2025} further reported detections of DMS and/or DMDS using MIRI, although several follow-up studies do not agree with their retrieval claims. \cite{Taylor2025} found that the MIRI data obtained by \cite{Madhusudhan2025} show no significant spectral features based on a flat-line rejection test. \cite{Welbanks2025} argued that Bayesian model comparisons with a small set of species do not directly translate into a statistically significant detection, and further demonstrated that several other hydrocarbons, such as propyne (\ce{C3H4}), can yield higher Bayesian evidence. \citet{Stevenson2025} again question the fidelity of the MIRI data from a single transit by highlighting the pixel-level systematics.

Following these studies, \citet{Hu2025} noted that including DMS in the base model does not significantly increase the Bayesian evidence (1.3 Bayes Factor), and replacing DMS with alternative molecules such as \ce{CH3SH} or \ce{N2O} provides comparably good fits. Independent analyses by other groups have also argued that the claimed DMS feature is not statistically significant and may instead be attributable to other simple hydrocarbons, such as \ce{C2H6} \citep{Luque2025} or \ce{C2H4} \citep{Stevenson2025}. 

To date, several key questions remain open about the theoretical understanding of temperate sub-Neptunes like K2-18 b. First, since the earlier water detection has been revised to methane, there is no physical explanation for the apparently low water abundance. A common hypothesis is that water vapor condenses below the altitudes probed by transmission spectroscopy, i.e., cold trapping. However, K2-18 b receives stellar irradiation only slightly higher than that of Earth ($S = 1368\ \mathrm{W m^{-2}} \approx 1.005,S_0$; \citealt{Benneke2019}). Previous 1D and 3D climate models consistently find either no water condensation \citep{Scheucher2020,Yu2021,Tsai2021b,Huang2024} or only trace condensation \citep{Charnay2021}, with water vapor remaining above $1\%$ in all cases. Second, while multiple independent studies \citep[e.g., ][]{Schmidt2025,Luque2025,Welbanks2025,Stevenson2025} have reanalyzed the spectral data and retrievals in \cite{Madhusudhan2023,Madhusudhan2025}, a physical modeling framework that tests the self-consistency of the climate and composition is still critically needed.
Finally, from an astrobiology perspective on interpreting observations, proposed potential abiotic pathways for DMS production \citep{Reed2024,Hu2025} introduce ambiguity in DMS as a biosignature gas.

This study is motivated to address the above unresolved issues with K2-18 b and temperate sub-Neptunes in general. The paper is structured as follows. Section~\ref{sec:methods} describes our climate and photochemical modeling framework, as well as the laboratory cross-section measurements. Section~\ref{sec:key} presents the results for K2-18 b, providing a physical explanation for the low water abundance, photochemical consistency tests, and exploration of the feasibility of abiotic DMS production. Section~\ref{sec:discussions} discusses the implications of our findings for the Hycean versus sub-Neptune interpretations, and we conclude in Section~\ref{sec:conclusions}.




\section{Methods}\label{sec:methods}
\subsection{Climate modeling}
We use the radiative transfer model HELIOS \citep{Malik2019a, Malik2019b} to compute the temperature profiles in a radiative-convective equilibrium. Our nominal model adopts a sub-Neptune scenario with a thick \ce{H2}-dominated atmosphere at 100$\times$ solar metallicity \citep{Asplund2009} (O/H = 4.8978$\times$10$^{-2}$, C/H = 2.6915$\times$10$^{-2}$, N/H = 6.7608$\times$10$^{-3}$, S/H = 1.3183$\times$10$^{-3}$, He/H = 0.08511). The assumed metallicity is consistent with the empirical mass–metallicity relation inferred from Solar System giants and exoplanets \citep{Welbanks2019}. We use GJ 436 (M2.5V) from the PHOENIX library \citep{Husser2013} as an analogous star, and a moderate intrinsic flux with $T_{\mbox{int}}$ = 60 K is adopted. The key update in this study is the inclusion of {\it radiative feedback from water clouds}, i.e., absorption and scattering by \ce{H2O}-ice particles with a prescribed size distribution. This treatment allows for quantifying the radiative impact of orbital eccentricity. 

For water clouds, we assume spherical particles with opacities computed from Mie theory. We assume a lognormal particle size distribution with a median radius of 10 $\mu$m and geometric standard deviation of 2, based on satellite measurements of marine liquid-water cloud particle sizes on Earth \citep{Han1994,Nakajima2001}. To obtain self-consistent radiative-photochemical equilibrium climate states, we iterate between the climate model HELIOS and the photochemical model VULCAN \citep{Tsai2017,Tsai2021}, until a steady state is reached. The active opacity sources included in the iteration are \ce{H2O}, \ce{CH4}, CO, \ce{CO2}, \ce{NH3}, \ce{C2H2}, and collision-induced absorption (CIA) from \ce{H2}-\ce{H2} and \ce{H2}-He (see Table 1 in \cite{Malik2019a} for the references of opacity sources). While the moist adiabat is used for the Hycean model, we assume a constant lapse rate of $R/c_p = 2/7$ (for diatomic ideal gas) in our sub-Neptune simulations with water clouds to ensure numerical stability. Nevertheless, the condensation region occupies only a narrow pressure range and is not expected to significantly affect the overall thermal structure. For cloud-free simulations, the convergence criterion is the relative change in global temperature of less than 1$\%$ between each climate-photochemistry iteration. When clouds are included, the feedback of water clouds introduces oscillation in the climate state. In such case, we instead relax the criterion and simply require the relative mean temperature difference to remain below 10$\%$ for a statistical steady state.

\subsection{Photochemical modeling}
We use the photochemical model VULCAN to compute the atmospheric composition under the effects of photochemistry, vertical mixing, and exchange with the surface or interior. The same 100$\times$ solar metallicity is adopted for the sub-Neptune models. While the strength of vertical mixing is difficult to constrain from first principles, typical choices span $K_{zz} = 10^3$ to $10^9$ cm$^2$ s$^{-1}$ or an Earth-like profile \citep{Hu2021,Wogan2024}. Since sub-Neptunes exhibit a climate dynamical regime distinct from that of Earth (e.g., slow rotation, a larger scale height, and potentially stationary stellar irradiation), we simply assume a vertically constant eddy diffusion coefficient of $K_{zz}$ = 10$^7$ cm$^2$s$^{-1}$. This value is broadly consistent with that derived from vertical wind speeds in general circulation models (GCMs) of K2-18 b \citep{Charnay2015,Tsai2021,Innes2022}. 

To investigate the diverse organic chemistry of sub-Neptunes, we adopt the S-N-C-H-O chemical network including DMS from \cite{Tsai2024b}\footnote{\url{https://github.com/shami-EEG/VULCAN/blob/master/thermo/SNCHO_DMS_photo_network_Tsai2024.txt}}. In addition, we further update the chemical network to include additional species and relevant reactions, as summarized below. First, we now treat the two isomers of \ce{C3H4} -- \ce{CH3CCH} (propyne) and \ce{CH2CCH2} (allene), as distinct species. Second, we add two additional sulfur-bearing species: \ce{H2CS} and \ce{C2H5SH}. \ce{H2CS} has been suggested as a key intermediate to form the sulfur-carbon bond in \ce{CH4}-rich environments \citep{Moses2024a}, while \ce{C2H5SH} is a structural isomer of DMS \citep{Calmonte2016,Hanni2024}. However, \ce{C2H5SH} remains in negligible abundance in the absence of chlorine and excited singlet sulfur (S$^1$D) \citep{Ghosh2025}. Lastly, because sulfur is not included in our climate modeling, once the self-consistent T-P profiles are obtained in Section~\ref{sec:apo}, we explore the atmospheric chemistry in Section~\ref{sec:dms} using the same T-P profile without re-evolving them. 

\subsection{Laboratory IR Cross-Section Measurements of DMS, DMDS, and DMSO}\label{sec:felix}
Motivated by the need for updated cross-section measurements for biogenic sulfur gases \citep{Schwieterman2018}, we conduct infrared spectroscopy measurements of DMS, DMDS, and DMSO using the FELIX free-electron laser at the HFML-FELIX institute\footnote{HFML-FELIX, Nijmegen 6525 ED, The Netherlands}. Full experimental details are presented in a companion paper \citep{Kuipers2026}; here we provide a brief summary. DMS, DMDS, and DMSO are expanded supersonically with Ar to form a cold molecular beam of neutral molecules at a temperature of approximately 40 K \citep{Lemmens2023}. The molecular beam is irradiated by the FELIX free-electron laser, inducing photodissociation when the photon energy is resonant with a vibrational transition. Infrared spectra can then be derived by comparing mass spectra recorded with and without infrared irradiation \citep{Jaeqx2016,Ferrari2024a}.

The experiments were conducted over the range of 650–3400 cm$^{-1}$ (2.95–15.4 $\mu$m) with a step size of 1 cm$^{-1}$. For comparison, the PNNL database provides measurements over a wider range, 600–6500 cm$^{-1}$ (1.54-16.6 $\mu$m), at higher spectral resolution (0.06 cm$^{-1}$). Our gas-phase experiments capture ro-vibrational lines without pressure-broadening effects, providing robust line positions and relative intensities. However, the caveat is that the absolute values of the cross sections cannot be determined due to the lack of knowledge of the sample number density. We therefore adopt a semi-experimental approach. After determining the line positions from the experiments, we apply density functional theory (DFT) calculations of the vibrational spectra to derive the absolute cross sections \citep{Ferrari2024}. The Doppler line widths are then determined from molecular dynamics simulations at 40, 100, 200, and 300 K, where the width is increased from 3.5 at 40 K to 6 cm$^{-1}$ at 300 K.



\begin{figure}[h]
    \centering
    \includegraphics[width=\linewidth]{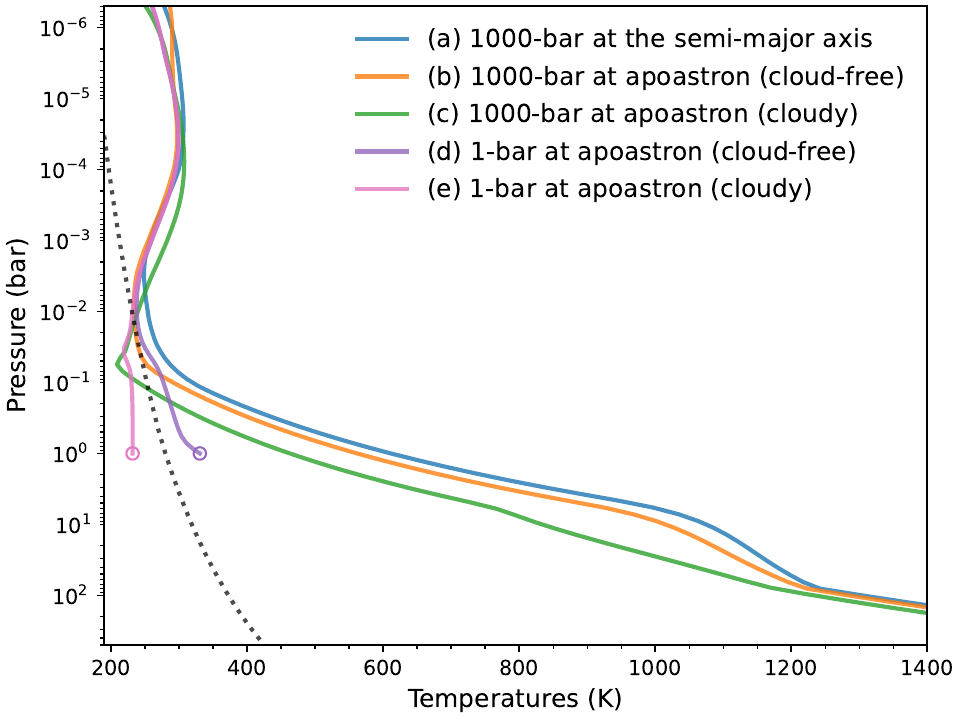}
    \caption{Temperature profiles of K2-18 b for different physical assumptions (see Section~\ref{sec:apo} for details):
    (a) Thick atmosphere with instellation received at the semi-major axis (S $\approx$ 1.005; \citealt{Benneke2019}).
    (b) Thick atmosphere with instellation received at apoastron, assuming e = 0.2 and no clouds.
    (c) Same as (b), but including \ce{H2O} clouds.
    (d) Thin atmosphere with a 1-bar surface, adopting the same instellation and composition from (b).
    (e) Same as (d), but including \ce{H2O} clouds. The condensation curve for \ce{H2O} for a mixing ratio of 0.01 is shown as a black dotted line to illustrate the location of the water cloud deck.}
    
    \label{fig:TPs} 
\end{figure}

\begin{figure}[h]
    \centering
    \includegraphics[width=\linewidth]{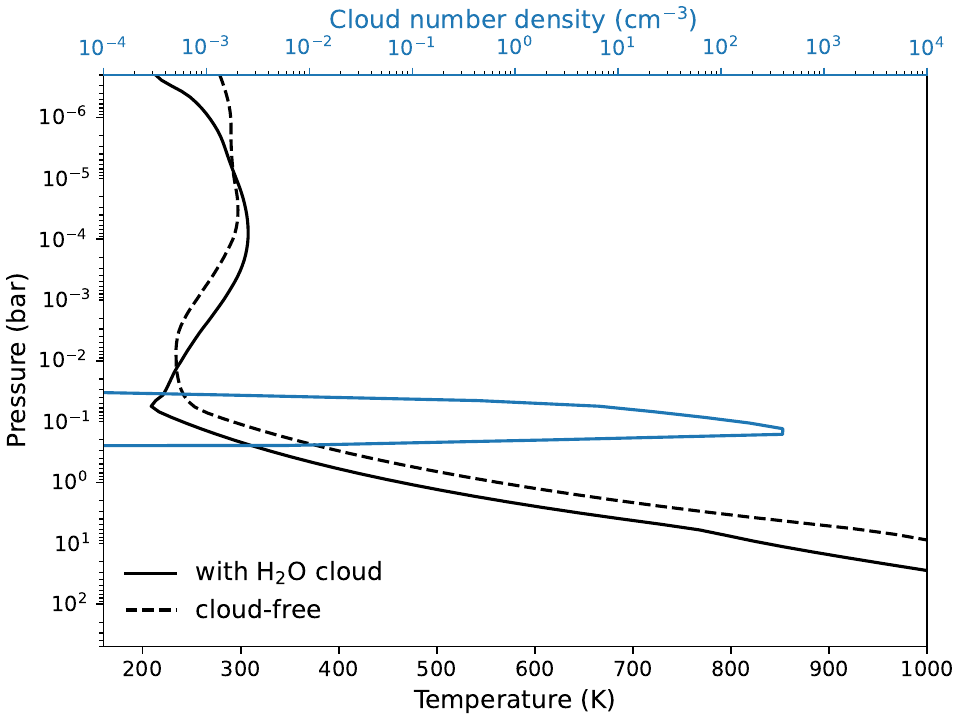}\\[-3pt]
    \includegraphics[width=1.075\linewidth]{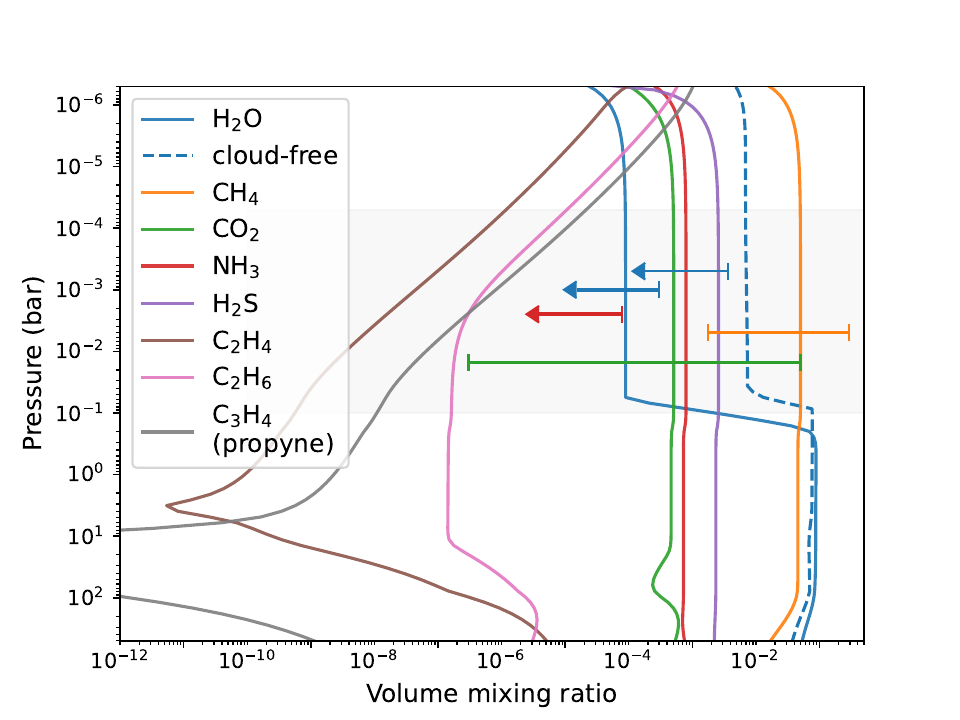}
    \caption{The temperature and cloud profiles (top) and gas volume mixing ratios (bottom) for K2-18 b at apoastron in a sub-Neptune scenario. In the top panel, the calculation including \ce{H2O} clouds is shown with a solid line, with the number density of clouds shown in blue. The temperature and \ce{H2O} abundance profiles for the cloud-free case are shown in dashed lines in both panels for comparison.
    In the bottom panel, the error bars and arrows represent the joined 1-$\sigma$ ranges and 2-$\sigma$ upper limits adopted from multiple retrieval analysis (Table \ref{tab:retrievals}). For \ce{H2O}, the thin line corresponds to \cite{Schmidt2025} and the thick line corresponds to \cite{Madhusudhan2023}, respectively.}
    \label{fig:apo}
\end{figure}

\section{Three key implications}\label{sec:key}

\subsection{A mildly eccentric orbit (e = 0.2) naturally explains the water cold trap}\label{sec:apo}
\ce{H2O} has not been robustly detected on K2-18 b since the original water claim from HST was overturned by JWST. Current spectral analyses of the NIRSpec data place a 2-$\sigma$ upper limit on the volume mixing ratio of \ce{H2O} of about 10$^{-3.5}$--10$^{-2.5}$ \citep{Madhusudhan2023,Schmidt2025,Luque2025}. Regardless of the interior and envelope structure of the planet, it is unlikely that water is intrinsically depleted on K2-18 b or other temperate sub-Neptunes. In the sub-Neptune scenario, enhanced metallicity predicts water abundance of 1--10$\%$, while in the Hycean scenario, water vapor should approach the saturation limit near the surface. 

To ensure efficient water condensation, previous theoretical studies have either 
imposed an ad hoc 30$\%$ reduction in the incident stellar flux \citep[e.g., ][]{Yang2024,Wogan2024,Tsai2024b,Hu2025,Huang2024} or applied a high Rayleigh enhancement factor \citep[e.g., ][]{Piette2020,Rigby2024} for temperate sub-Neptunes. We note that this approach cannot be justified by the planetary energy budget and likely stems from a misconception of Earth's Bond albedo. On Earth, low-level (below $\sim$2 km) clouds account for roughly half of the reflected shortwave radiation \citep[e.g., ][]{Goldblatt2011,Stephens2012,Tang2020}, while clear-sky Rayleigh scattering and surface reflection contribute the remainder \citep[e.g., ][]{Kitzmann2010,Donohoe2011}. A simple treatment in 1D climate models is to parametrize the combined albedo effects of atmosphere, clouds, and surface by enhancing the effective surface albedo \citep{Fauchez2018,Charnay2021}. This parameterization is physically different from reducing the stellar flux at the top of the atmosphere. We highlight two major pitfalls of imposing reduced irradiation at the top of the atmosphere: (1) it artificially removes energy that would otherwise be absorbed by the atmosphere and surface \citep{Jordan2025}; and (2) when applied in a numerical radiative-transfer calculation, it double-counts the clear-sky Rayleigh scattering by gas molecules. 


Here, we propose that the mild orbital eccentricity of K2-18 b provides a physically motivated explanation for water rainout. High-resolution RV measurements with CARMENES and HARPS \citep{Sarkis2018} yield an orbital eccentricity of e = 0.20 $\pm$ 0.08. At apoastron, the instellation decreases to
\begin{equation}
S_a = \frac{S_e}{(1+e)^2},
\end{equation}
where $S_e$ is the instellation on a circular orbit or at the semi-major axis. For e = 0.2, $Sa \sim 0.69 S_e \sim 0.7 S_0$, corresponding to $\sim 30\%$ reduction in stellar flux.

Figure~\ref{fig:TPs} illustrates the temperature profiles considering different planetary scenarios and cloud conditions. Profiles (a) and (b) are our self-consistent, cloud-free sub-Neptune models in radiative-convective equilibrium for a circular orbit and at apoastron for an eccentric orbit, respectively. As the instellation decreases by $\sim 30\%$ at apoastron, temperatures in the infrared photosphere ($\sim$1 bar--1 mbar) decrease by 10--30 K. Profile (c) includes the radiative feedback from cloud scattering. Below the cloud top ($P \gtrsim 0.05$~bar), scattering by the clouds lowers the photospheric temperature by $\sim$50--100~K relative to the cloud-free case. Above the cloud deck ($\sim$10$^{-2}$--10$^{-4}$~bar), the stellar radiation reflected by the clouds is partly absorbed, mainly by \ce{CH4}, which warms these layers by $\sim$20~K. We discuss this cloud--climate feedback further in Section~\ref{sec:cloud-feedback}.


For a direct comparison with the habitable Hycean scenario with a thin atmosphere, profiles (d) and (e) adopt the same atmospheric composition as model (b), but truncated at 1 bar. These thin atmospheres differ from those in \cite{Tsai2024b}, where the atmospheric composition was self-consistently calculated for a 1-bar atmosphere, resulting in lower greenhouse gas abundances (e.g., \ce{CO2}, \ce{CH4}, and \ce{H2O}) and consequently lower temperatures. Here, profile (d) shows that in our 1-bar, cloud-free atmosphere at apoastron, the surface temperature is about 330 K. When water clouds are included in profile (e), the near-surface layers become more isothermal, further lowering the surface temperature to about 230 K, below the freezing point of water.

The effects of \ce{H2O} clouds on a temperate sub-Neptune are further illustrated in Figure~\ref{fig:apo}. Under the same stellar irradiation at apoastron, the \ce{H2O} cloud deck near the top of the convective region strengthens the water cold trap, reducing the \ce{H2O} vapor mixing ratio from about 1$\%$ to 10$^{-4}$. Consequently, only when both the reduced stellar irradiation and the radiative effects from \ce{H2O} clouds at apoastron are taken into account does our climate–photochemical model yield a water abundance consistent with the observational upper limits. We examine the cloud–climate feedback in more detail in the following section and discuss the overall atmospheric composition in Section~\ref{sec:dms}.



\begin{figure}[h]
    \centering
    \includegraphics[width=\linewidth]{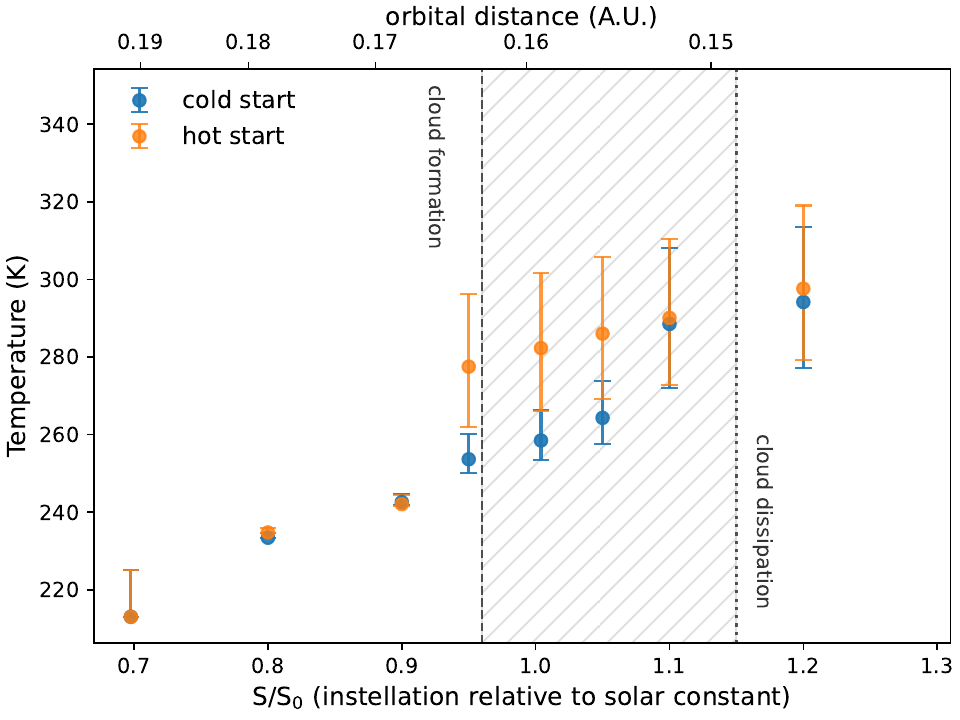}
    \caption{The temperatures around water's cold trap as a function of stellar irradiation normalized by Earth's insolation. Simulations initialized from a cold state (from left to right) are shown in blue, while those initialized from a hot state (from right to left) are shown in orange. Circles indicate temperatures at 50 mbar, and the error bars represent the temperature range between 25 and 75 mbar, chosen to be near the cloud top region. The dashed line marks the onset of spontaneous water condensation and cloud formation, whereas the dotted line indicates the evaporation and dissipation of water clouds. The slash-line shaded region enclosed by the dashed and dotted lines corresponds to the bifurcation regime where two equilibrium solutions can exist.}
    \label{fig:bifur}
\end{figure}

\begin{figure}[h]
    \centering
    \includegraphics[width=\linewidth]{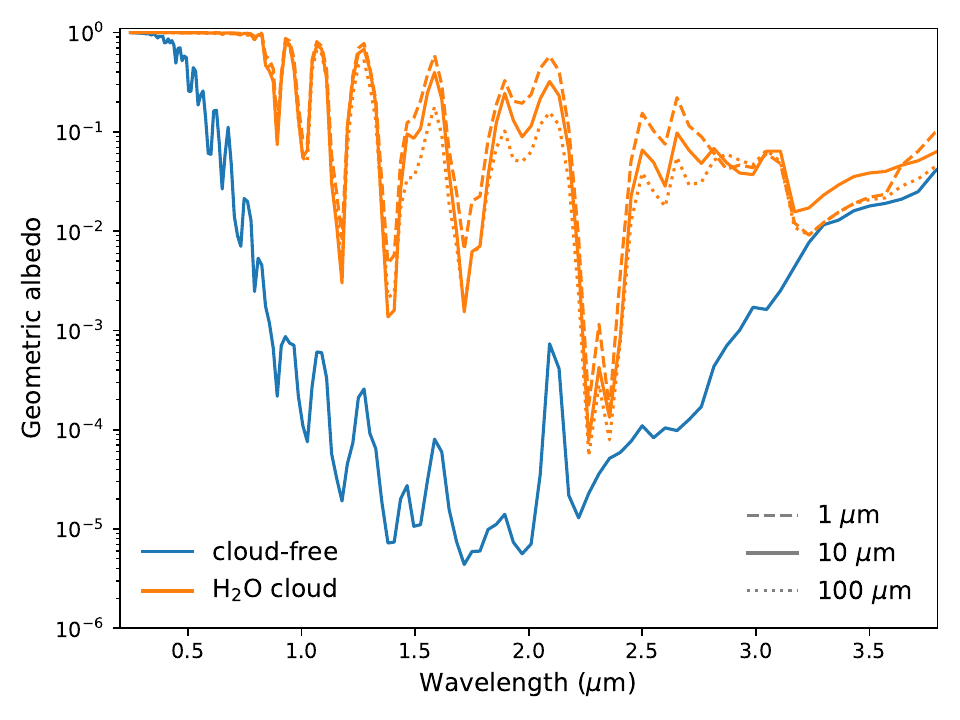}
    \includegraphics[width=\linewidth]{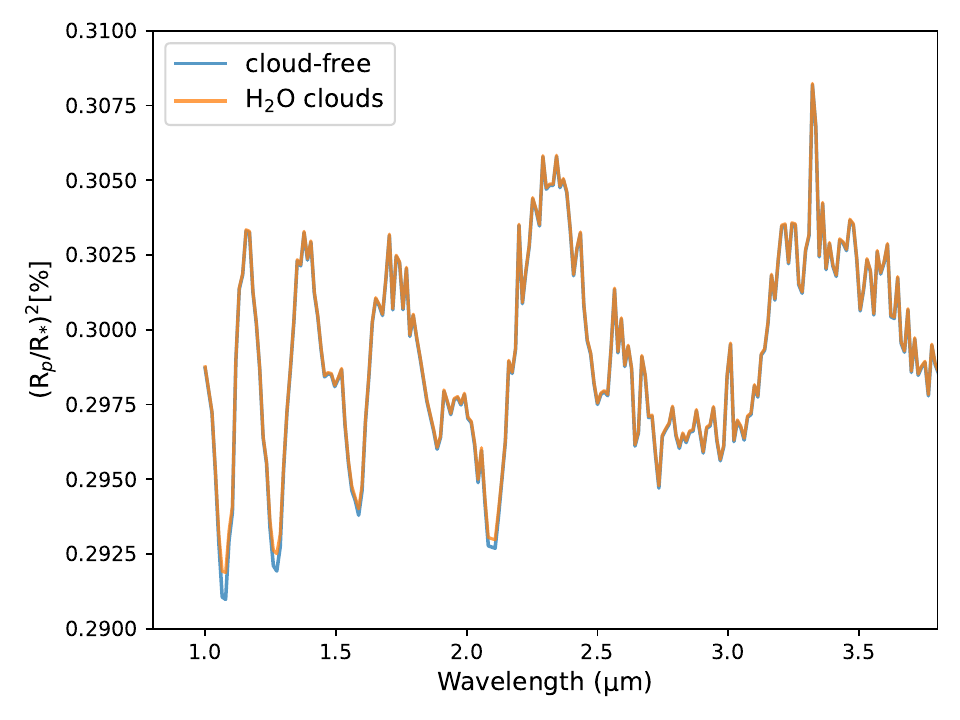}
    \caption{Top: Geometric albedo of the K2-18 b atmospheric model at apoastron, for the cloud-free case (blue) and with \ce{H2O} clouds (orange) of medium particle radii 1~$\mu$m (dashed), 10~$\mu$m (solid; nominal case), and 100~$\mu$m (dotted). Bottom: Corresponding transmission spectra for the cloud-free model (blue) and for a model with a gray cloud deck at a cloud-top pressure of 0.02 bar (orange).}
    \label{fig:geo_albedo}
\end{figure}

\subsubsection{Climate bifurcation due to clouds}\label{sec:cloud-feedback}
To demonstrate the radiative feedback of \ce{H2O} clouds, we run a series of 1D climate models with varying initial conditions using HELIOS. Beginning at periastron, we progressively decrease the instellation, using the radiative–convective equilibrium temperature from the previous step to initialize the subsequent run, i.e., ``hot--start" sequence. We then repeat the sequence beginning at apoastron, i.e., ``cold--start" sequence. The resulting temperatures around the cloud-forming region from the two sequences are summarized in Figure~\ref{fig:bifur}. We find that over the instellation range  $0.95S_0 \lesssim S \lesssim 1.05S_0$, the cloudiness state depends on the initial condition: initially cloud-free atmospheres remain cloud-free, while cloudy atmospheres remain cloudy. This behavior indicates the two stable radiative-equilibrium states within this bifurcation regime.      

The impact of water cloud optical properties on the geometric albedo spectrum is illustrated in Figure~\ref{fig:geo_albedo}. Scattering dominates in the optical and near-infrared up to $\sim$ 3 $\mu$m, while absorption in the longer wavelengths increases with particle size as it enters the Mie regime. For the same condensed water mass, smaller particles have a larger total surface area and are expected to produce a higher albedo. However, the geometric albedo depends only weakly on the particle size distribution (Figure~\ref{fig:geo_albedo}, top panel). Once clouds form, their scattering on the stellar flux reduces temperatures near the cloud level while increasing temperatures above it. The bottom panel of Figure~\ref{fig:geo_albedo} also demonstrates that these low-level clouds do not significantly impact the transmission observations.

When the radiative effects of water clouds are included, cold trapping becomes more efficient, substantially lowering the atmospheric water abundance. Although cloud radiative feedback is intrinsically three-dimensional and the quantitative cloud distribution should be assessed with 3D GCMs \citep[e.g., ][]{Yang2013a,Charnay2015,Turbet2021}, our simplified 1D framework nonetheless captures the qualitative bifurcation behavior. In our 1D model, the net effect of deep water clouds is to increase the shortwave ($\lesssim$ 3$\mu$m) albedo and cool the infrared-emitting levels. Once clouds form, the resulting tropospheric cooling promotes further condensation, establishing a positive feedback that reinforces cloud formation.









\subsection{Methane does not appear in isolation in a Hycean scenario}\label{sec:dms} 
Given the current observational constraints based on the JWST NIRSpec and MIRI data \citep{Madhusudhan2023,Madhusudhan2025,Hu2025} and the subsequent independent Bayesian retrievals \citep{Schmidt2025,Luque2025,Stevenson2025}, one critical question arises --  are the inferred atmospheric abundances mutually consistent? Photochemical modeling can provide insights into this question. Essentially, we will assess whether the detected gases are expected to coexist under assumed atmospheric conditions, and conversely, whether additional gases are expected to coexist with them but are undetected.

Table~\ref{tab:retrievals} summarizes the current abundance constraints reported in recent spectral-retrieval studies. We highlight the main results as follows. All studies consistently report a detection of \ce{CH4}, with retrieved abundances of order 1 -- 10 $\%$. For \ce{CO2}, although its feature was not found to be statistically significant in the NIRISS and NIRSpec data reanalysis of \citet{Madhusudhan2023} by \citet{Schmidt2025}, it was subsequently confirmed when follow-up observations with the NIRSpec G235H and G395H modes were included \citep{Hu2025}. The non-detections of important molecules include \ce{H2O} and \ce{NH3}, with 2-$\sigma$ upper limits of approximately 10$^{-3}$ and 10$^{-4}$, respectively. Finally, DMS and DMDS have so far been confidently reported only in the MIRI analysis by \citet{Madhusudhan2025} and tentatively reported by \citet{Madhusudhan2023,Hu2025}. Subsequent analyses of the MIRI data found no evidence for DMS or DMDS \citep{Taylor2025,Welbanks2025}.

We now examine compositional compatibility by adopting the reported abundance constraints of the detected species at face value. We impose these constraints in the photochemical model to test whether the reported detections and non-detections can be realized in chemically consistent steady states. We apply this approach to both the Hycean and sub-Neptune scenarios in Sections~\ref{sec:hycean} and \ref{sec:nep}.

\begin{table*}[!t]
\centering
\begin{tabular}{l|c|c|c|c|c|c|c} 
Ref. & M23\footnote{Table 2 in \cite{Madhusudhan2023} across all three cases} & M25\footnote{Tables 2 and 3 in \cite{Madhusudhan2025} including DMS-only and DMDS-only cases in the canonical model}& Sc25 \footnote{Quoting the median 2$\sigma$ upper limits for all species but \ce{CH4} from \cite{Schmidt2025}} & L25 \footnote{From fitting models with DMS/DMDS/\ce{C2H6} to JExoRES and exoTEDRF+SPARTA datasets from Table C.1. in \cite{Luque2025}} & PC25 \footnote{Table 3 with MIRI datasets in \cite{PicaCiamarra2025}}& Hu25\footnote{Tables 5 and 6 in \cite{Hu2025}} & St25\footnote{Adopting the widest range across MIRI pixel shifts 5, 6, and 7 in \cite{Stevenson2025}}\\ 
Wavelength range & 0.8--5.2 $\mu$m & 4.8--12 $\mu$m & 0.8--5.2 $\mu$m & 0.6--12 $\mu$m & 0.8--12 $\mu$m & 0.9--5.2 $\mu$m & 0.8--12 $\mu$m\\
(Instruments) & (I, II) & (III) & (I,II) &  (I,II,III)   &  (I,II, III) &   (II,IV) & (I,II, III) \\
     \hline       
\ce{CH4} & -2.76 -- -1.15 & -- & -2.14 -- -0.53 & -1.61 -- -0.58 & -- &  -1.61 -- -0.83& -3.04 -- -0.9\\ 
\ce{CO2} & -3.03 -- -1.30 & -- & $<$ -1.58 & -6.52  -- -1.42  & -- & -4.91 -- -2.04& -8.13 -- -1.38\\ 
\ce{H2O} &  $<$ -3.06 & -- &  $<$ -2.44 &  $<$ -2.72  & -- & $<$ -3.52 & $<$ -1.13\\ 
\ce{NH3} & $<$ -4.46 & -- &  $<$ -4.70 &  $<$ -4.11  & -- & $<$ -4.93 & $<$ -2.09\\ 
DMS & -4.46 -- -3.69 & -4.86 -- -2.26 & $<$ -3.58 & $<$ -2.79 & -5.2 -- -2.61 & -7.22 -- -4.92 & $<$ -1.78\\
DMDS & -- & -4.75 -- -2.08 & -- & $<$ -4.12 & -4.60 -- -2.34 & -- & $<$ -2.4 \\  
\ce{CS2} &  --     & -- & $<$ -2.63 & -- & -- & -- & $<$ -1.68\\
\ce{C2H4} &  -- & -- & -- & -- & -- & -- & -7.62 -- -1.23\\ 
\ce{C2H6} &  --     & -- &  -- & $<$ -1.05 & -- & --  &\\
\ce{C3H4} (propyne) &  --  & -- & -- & -- & -4.21 -- -1.86 & -- \\ 
\end{tabular}
\caption{A compilation of retrieved volume mixing ratios, listed in log$_{10}$, from recent studies. The wavelength ranges and included instruments are noted as I: NIRISS SOSS, II: NIRSpec G395H, III: MIRI, IV: NIRSpec G235H. From each retrieval analysis, we adopt the conservative range of the lowest -1$\sigma$ and the highest +1$\sigma$ values for constrained abundance and the highest value for the 2$\sigma$ upper limits, unless stated otherwise.
}
\label{tab:retrievals}
\end{table*}

\subsubsection{Hycean scenario}\label{sec:hycean}

We adopt the thin, cloudy atmosphere model at apoastron (profile (e) in Figure~\ref{fig:TPs}) to represent a proposed habitable planet with surface oceans. \ce{H2O} is set to the saturation mixing ratio at the surface with 80$\%$ relative humidity. To explore the observational implications of such a Hycean planet under the reported composition constraints, we impose fixed abundances of \ce{CH4} and \ce{CO2} as lower boundary conditions at the surface to match the retrieval constraints. This is necessary because, without fixed boundary constraints, \ce{CH4} would be lost through photochemical processes \citep{Tsai2021,Wogan2024}. Guided by multiple retrieval analyses, we adopt [\ce{CH4}] = 5$\times$10$^{-2}$ and [\ce{CO2}] = 5$\times$10$^{-4}$ at the surface. 

In addition, we prescribe a fixed distribution of DMS to further assess its plausibility under the claimed detection scenario. Because DMS is highly susceptible to photolysis, we fix its mixing ratio to 1$\times$10$^{-4}$ from the surface up to 10$^{-3}$ bar in order to maintain the abundance in the region probed by transmission spectroscopy. We note that this experiment differs from the self-consistent modeling in \cite{Tsai2024b}, where organic sulfur species are supplied by prescribed biological fluxes at the surface and \ce{CH4} is allowed to be photochemically destroyed throughout the atmosphere.

\begin{figure*}[t]
    \centering
    \includegraphics[width=0.75\linewidth]{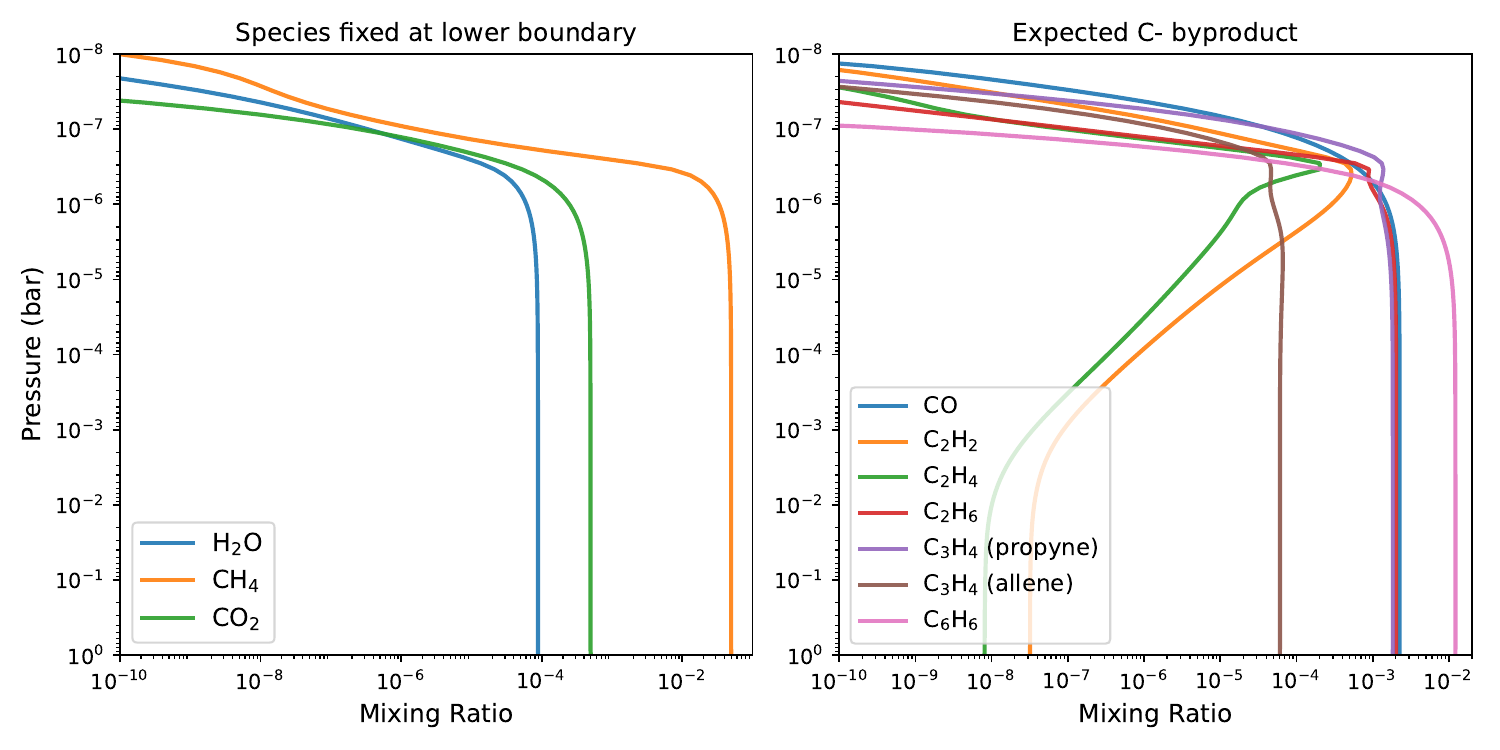}
    \caption{The atmospheric compositions of our cloudy Hycean model based on the temperature profile (e) in Figure~\ref{fig:TPs}. The left panel shows the species whose abundances are fixed at the surface (see Section \ref{sec:hycean}). The right panel highlights the abundant carbon-bearing molecules under the assumed fixed abundances and Hycean conditions.}
    \label{fig:hycean-fixed-ch4}
\end{figure*}

Figure~\ref{fig:hycean-fixed-ch4} presents the results of our Hycean model where \ce{CH4} and \ce{CO2} are fixed at values consistent with observational constraints. The elevated \ce{CH4} abundance within a thin atmosphere with a relatively dry stratosphere leads to highly efficient hydrocarbon production, consistent with the previous study by \cite{Huang2024}. Benzene (\ce{C6H6}) reaches 1$\%$, followed by \ce{C2H6} and \ce{C3H4} at levels just above $10^{-3}$. 

The formation of \ce{C2H6} has been extensively studied in the context of Jupiter, where it proceeds primarily through methyl (\ce{CH3}) recombination \citep{Moses2005}. We find the same dominant pathway, i.e.
\begin{eqnarray}
\begin{aligned}
\ce{CH4 + H &-> CH3 + H2}\\
\ce{CH3 + CH3 &->[M] C2H6}\\
\noalign{\vglue 3pt}
\end{aligned}
\label{re:CH4=C2H6}
\end{eqnarray}
Once ethane becomes abundant, \ce{C2H6} photolysis initiates the production of \ce{C3H4} (propyne) through  
\begin{eqnarray}
\begin{aligned}
\ce{CH4 + H &-> CH3 + H2}\\
\ce{CH3 + CH3 &->[M] C2H6}\\
\ce{C2H6 &->[h\nu] C2H2 + 2H2}\\
\ce{CH3 &->[h\nu] ^1CH2 + H}\\
\ce{^1CH2 + H2 &-> CH2 + H2}\\
\ce{CH2 + C2H2 &->[M] C3H4 (CH3CCH)}\\
\noalign{\vglue 3pt}
\end{aligned}
\label{re:CH4=C2H6}
\end{eqnarray}

We note that benzene should be interpreted as a precursor to organic hazes, as its accumulation is likely an artifact of truncating the chemical network at six-carbon molecules. In reality, the formation of the first aromatic ring would conceivably initiate further polymerization, leading to the growth of more complex hydrocarbons and photochemical hazes \citep{He2018,Horst2018,Vuitton2019}. Such hydrocarbon hazes have been suggested by the recent retrieval analysis of \cite{Liu2026}. Our prediction of efficient haze-precursor production in the Hycean scenario is consistent with \cite{Wogan2024}, who found the greatest photochemical haze production in their methane-producing Hycean model. Unlike the hazy sub-Neptune LP~791-18~c \citep{Roy2026}, however, K2-18~b shows no conclusive haze-induced scattering slope. While \cite{Madhusudhan2023} found that enhanced Rayleigh scattering is generally favored, \cite{Schmidt2025} found no clear evidence for hazes. \cite{Jaziri2025} further proposed that the absorption of photochemical hazes could help reconcile the mismatch in spectral amplitude between the MIRI and NIRISS SOSS data, but the inferred hazes are sensitive to the assumed optical properties of haze analogs and instrumental offsets.

\begin{figure*}[t]
    \centering
    \includegraphics[width=\linewidth]{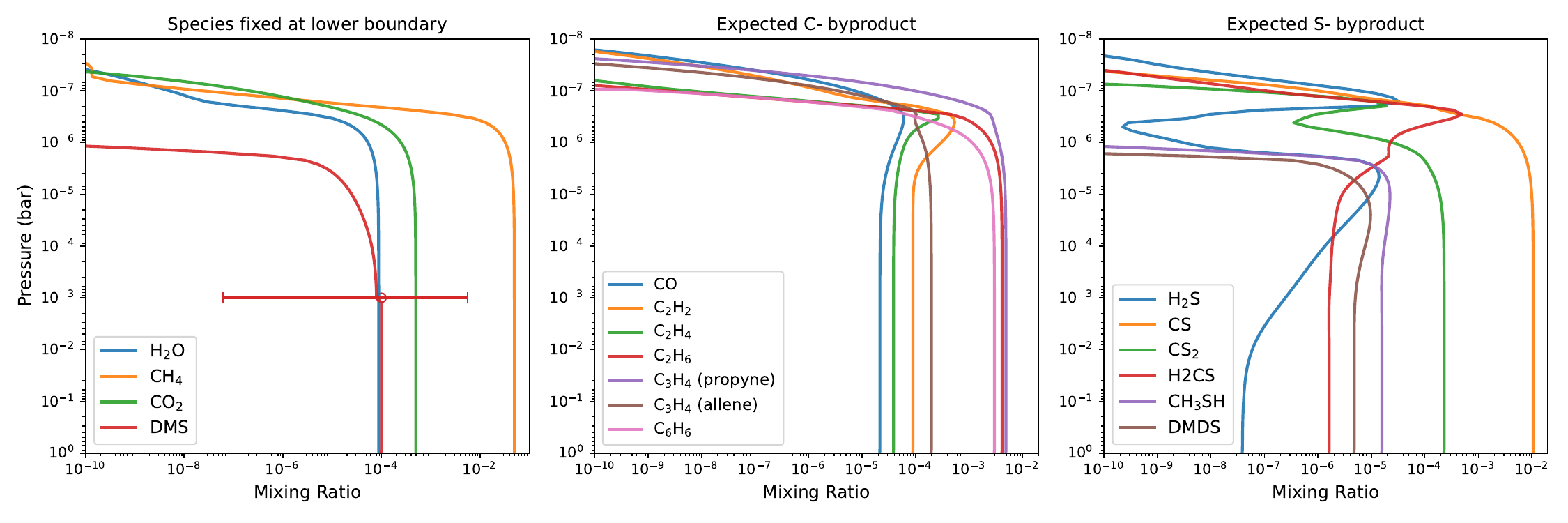}
    \caption{Same as Figure~\ref{fig:hycean-fixed-ch4}, but with DMS fixed at pressures between 1 and $10^{-3}$ bar. Sulfur byproducts are additionally shown in the rightmost panel.}
    \label{fig:hycean-fixed-dms}
\end{figure*}

Figure~\ref{fig:hycean-fixed-dms} shows that, in our model with imposed DMS at 100 ppm, \ce{C2H6}, \ce{C3H4}, and \ce{C6H6} remain the dominant carbon species at high abundances. The supply of DMS drives the atmosphere toward more reducing conditions, decreasing CO abundance while enhancing the production of unsaturated hydrocarbons such as \ce{C2H2} and \ce{C2H4}, consistent with the findings in \citet{Tsai2024b}. Sulfur also readily bonds with carbon to form CS and \ce{CS2}, which reach abundances of approximately 1$\%$ and 10$^{-4}$, respectively, exceeding those of DMDS and \ce{CH3SH}. The primary production pathway for CS and \ce{CS2} begins with the photodissociation of DMS, which produces \ce{CH3S} radical, and recombines to form both \ce{CH3SH} and \ce{H2CS}. The relatively weak C--H bond in \ce{H2CS} enables efficient onward production of CS and \ce{CS2} (see \cite{Moses2024a} for detailed mechanisms):

\begin{eqnarray}
\begin{aligned}
\ce{DMS &->[h\nu] CH3S + CH3}\\
\ce{CH3S + CH3S &-> CH3SH + H2CS}\\
\ce{H2CS &->[h\nu] H2 + CS}\\
\ce{CS + SH &-> CS2 + H}\\
\noalign{\vglue 3pt}
\end{aligned}
\label{re:DMS-CS2}
\end{eqnarray}

\begin{figure}[t]
    \centering
    \includegraphics[width=\linewidth]{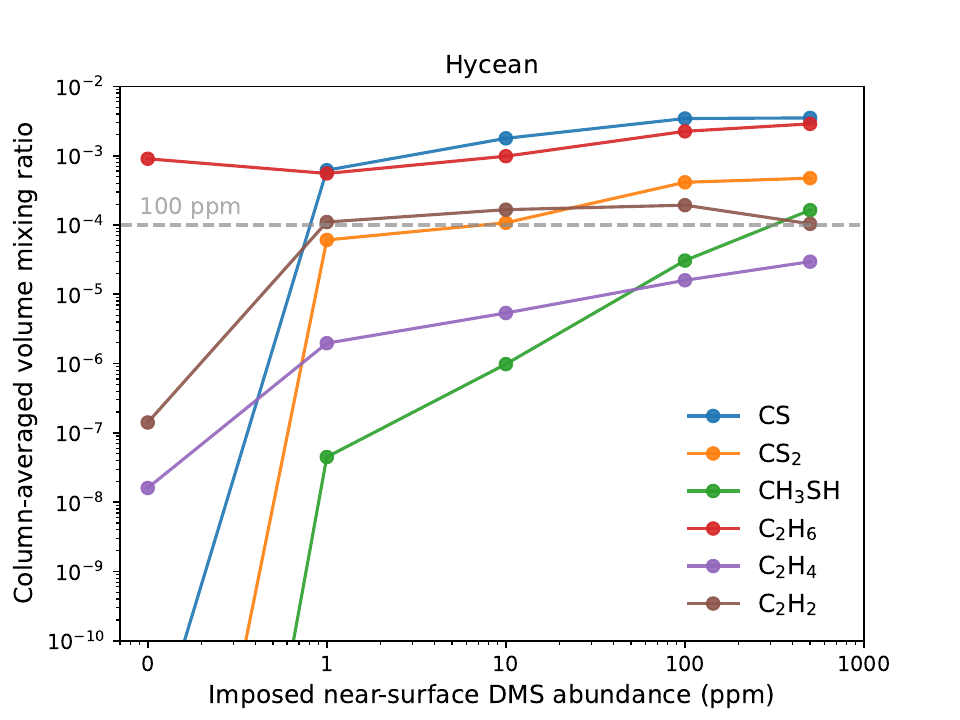}
    \includegraphics[width=\linewidth]{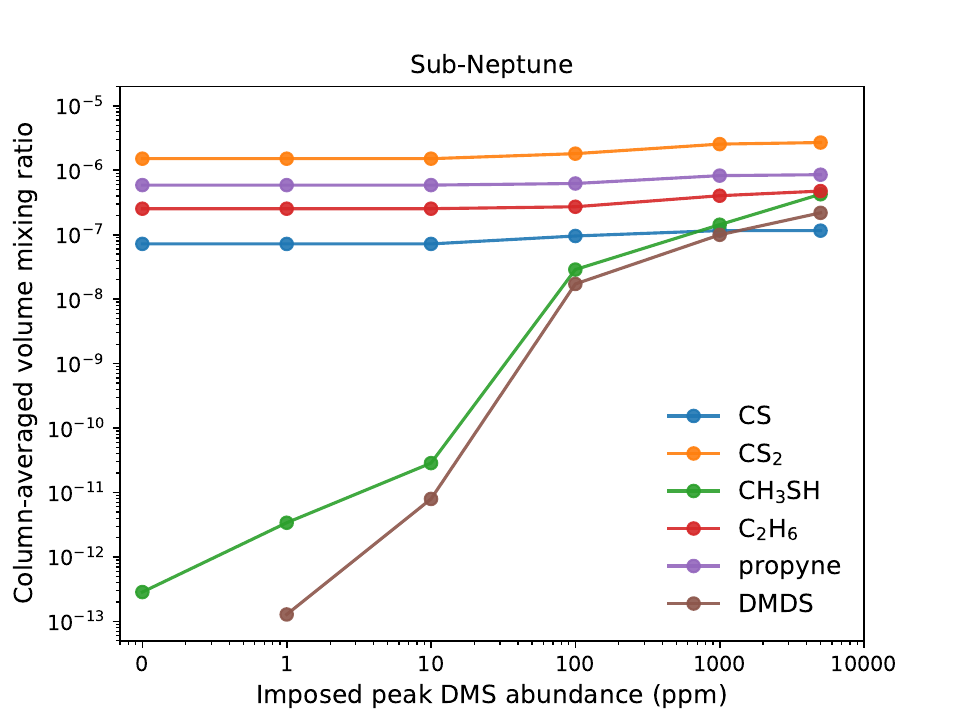}
    \caption{Response of carbon and sulfur byproducts to varying imposed DMS abundances in the Hycean (top) and sub-Neptune (bottom) scenarios. The imposed near-surface (from 1 to 10$^{-3}$ bar) DMS abundance varies from 0 to 500 ppm in the Hycean scenario, and peak abundance of the imposed DMS profile varies from 0 to 5000 ppm in the sub-Neptune scenario. The dashed line in the top panel marks the 100~ppm detectability threshold for \ce{CS2}. Note the different species plotted in the two panels.} 
    \label{fig:sen-dms}
\end{figure}

To aid the interpretation of observations, we further test the sensitivity of the atmospheric composition to the imposed near-surface DMS abundance, which varies from 0 (DMS-free) to 500~ppm. Figure~\ref{fig:sen-dms} shows the resulting abundances of the main hydrocarbon and sulfur byproducts. The tests show that \ce{CS2} and \ce{C2H2} can serve as complementary indicators of DMS in a Hycean atmosphere. Assuming a detectability threshold of $\sim$100~ppm for \ce{CS2}, based on the lowest \ce{CS2} abundance retrieved to date \citep{Zhang2026}, we find that both \ce{CS2} and \ce{C2H2} reach this level only when the near-surface DMS abundance is $\gtrsim 10$~ppm. In contrast, the DMS-free Hycean scenario has a \ce{C2H2} abundance below 1~ppm and negligible \ce{CS2}. \ce{CS2} and \ce{C2H2} can therefore discriminate between Hycean atmospheres with and without DMS.

Previous studies have demonstrated that \ce{CH4} is unstable against photolysis in a Hycean atmosphere and therefore requires active replenishment \citep{Wogan2024,Tsai2024b,Huang2024}. Our results in this study indicate that \ce{CH4} in a Hycean atmosphere leads to substantial hydrocarbon production, with species such as \ce{C2H6} and \ce{C3H4} reaching abundances of approximately 5$\%$ that of \ce{CH4}. If DMS is also present in a Hycean atmosphere, the production of unsaturated hydrocarbons is further elevated to potentially detectable levels. In this case, CS and \ce{CS2} are expected to emerge as the dominant photochemical products with strong infrared absorption features, exceeding the abundances of \ce{CH3SH} and DMDS. 

\begin{figure*}[t]
    \centering
    \includegraphics[width=0.85\linewidth]{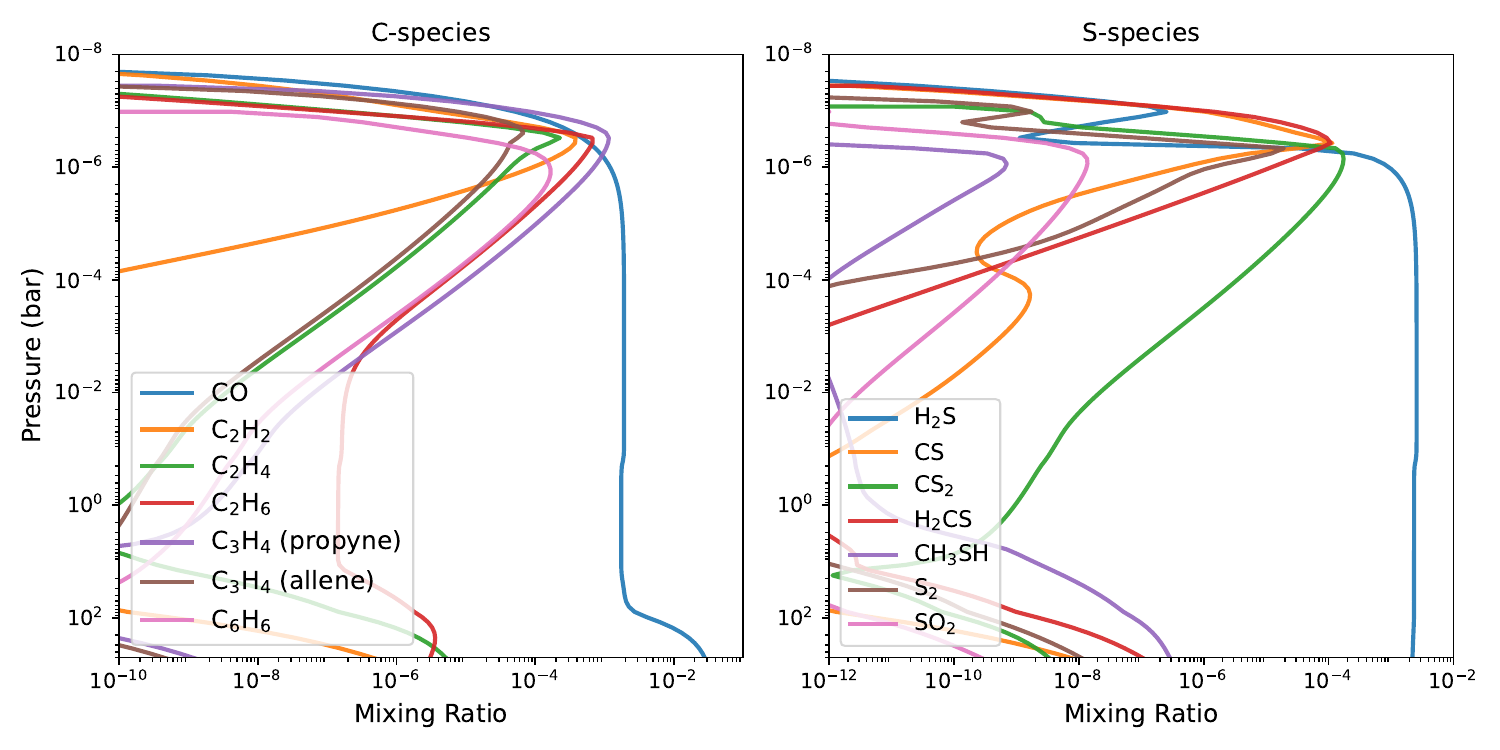}
    \caption{The atmospheric compositions of our cloudy sub-Neptune model based on the temperature profile (c) in Figure~\ref{fig:TPs}, the same model as in Figure~\ref{fig:apo}. 100$\times$ solar metallicity is assumed (without fixing any boundary conditions). The left panel highlights the carbon-bearing molecules while the right panel highlights the sulfur-bearing molecules.}
    \label{fig:nep}
\end{figure*}

\begin{figure*}[t]
    \centering
    \includegraphics[width=0.85\linewidth]{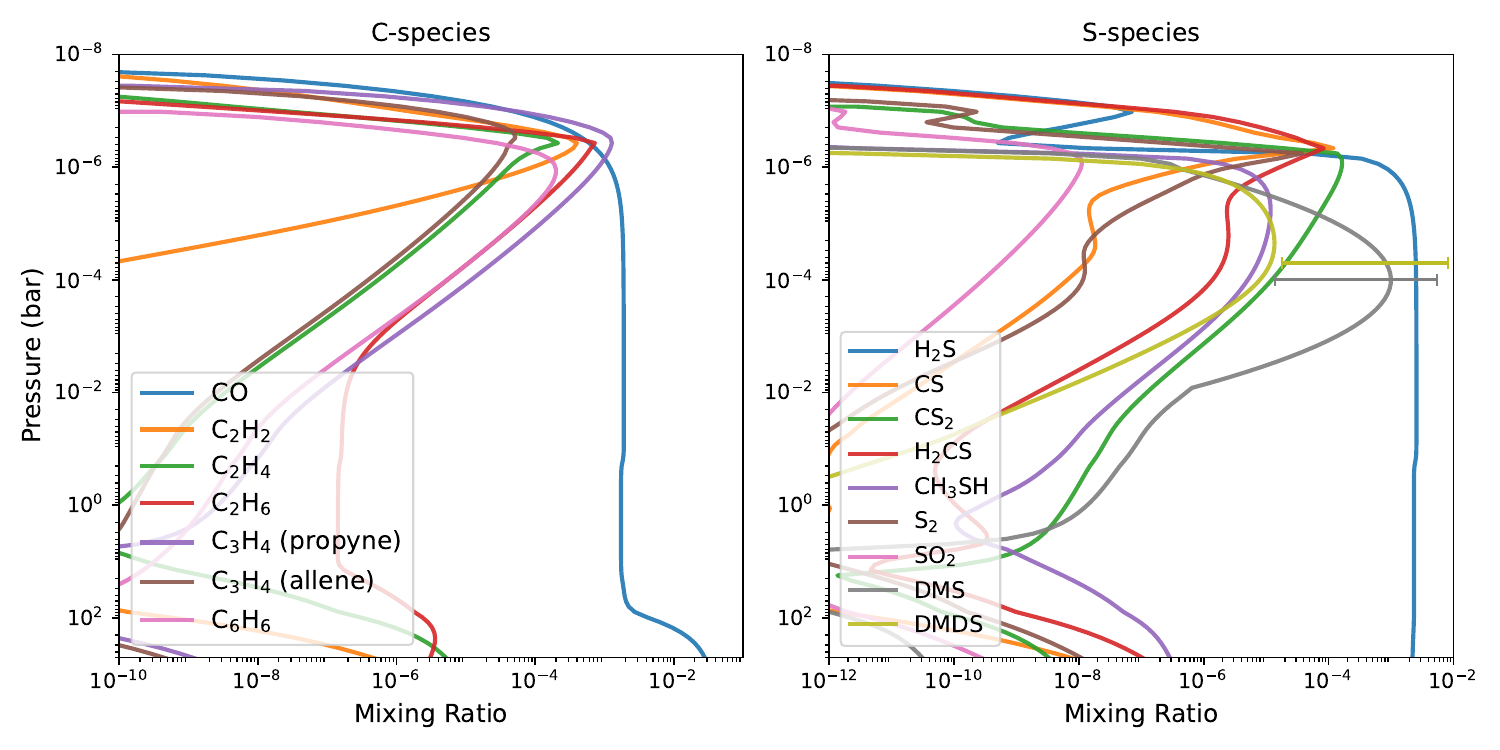}
    \caption{Same as Figure~\ref{fig:nep}, but with the DMS vertical distribution prescribed as a fixed Gaussian profile. The error bars in the right panel denote the 1-$\sigma$ range of the DMS and DMDS volume mixing ratios reported by \cite{Madhusudhan2025}.}
    \label{fig:nep-dms}
\end{figure*}

\subsubsection{Sub-Neptune scenario}\label{sec:nep}
Following the argument in Section~\ref{sec:apo}, we adopt the apoastron temperature profile with \ce{H2O} clouds for our sub-Neptune model, in which the water cold trap is sufficient to reduce \ce{H2O} to levels consistent with observational constraints. In this scenario, with a thick and high-metallicity (100$\times$ solar) \ce{H2}-atmosphere, \ce{CH4} and \ce{CO2} are maintained by equilibrium and atmospheric mixing processes, as shown in the bottom panel of Figure~\ref{fig:apo}. Figure~\ref{fig:nep} shows that hydrocarbons are also produced in the upper atmosphere near 10$^{-6}$ bar, but their abundances decline rapidly with increasing pressure as they are transported downward and destroyed in the deep layers. The key difference between the Hycean and sub-Neptune models is that hydrocarbons in the sub-Neptune case are confined to the upper atmosphere and at lower abundances. For instance, \ce{C2H6} (ethane) and \ce{C3H4} (propyne), the most abundant hydrocarbons, remain well below 1$\%$ of the \ce{CH4} abundance at pressures greater than $10^{-5}$ bar.

Similar to the test for the Hycean scenario, we next explore the case where DMS is present in a sub-Neptune atmosphere. Regardless of its origin, whether biogenic or abiotic, we prescribe a fixed DMS abundance profile following a Gaussian distribution in log-pressure, centered at 0.1 mbar with a peak abundance of 10$^{-3}$ and width of 0.5~dex, as shown in Figure~\ref{fig:nep-dms}. The imposed DMS leads to the production of \ce{CH3SH} and DMDS via the recombination of \ce{CH3S}, which is produced by DMS photolysis.   

Figure~\ref{fig:sen-dms} (bottom) also shows the sensitivity to the imposed Gaussian DMS profile in the sub-Neptune scenario. \ce{CS2} and propyne remain at $\sim$1~ppm and are nearly independent of the imposed DMS. \ce{CH3SH} and DMDS increase by several orders of magnitude with increasing DMS, but generally remain below 1~ppm. Apart from \ce{CH3SH} and DMDS, the additional DMS does not significantly alter the abundances of other atmospheric constituents. Thus, in a sub-Neptune where the deeper layers efficiently destroy organic sulfur species, the dominant photochemical byproducts of photospheric DMS are limited to \ce{CH3SH} and DMDS.

\subsubsection{Compositional consistency in Hycean and sub-Neptune scenarios under observational constraints}
\begin{figure*}[t]
    \centering
    \includegraphics[width=0.95\linewidth]{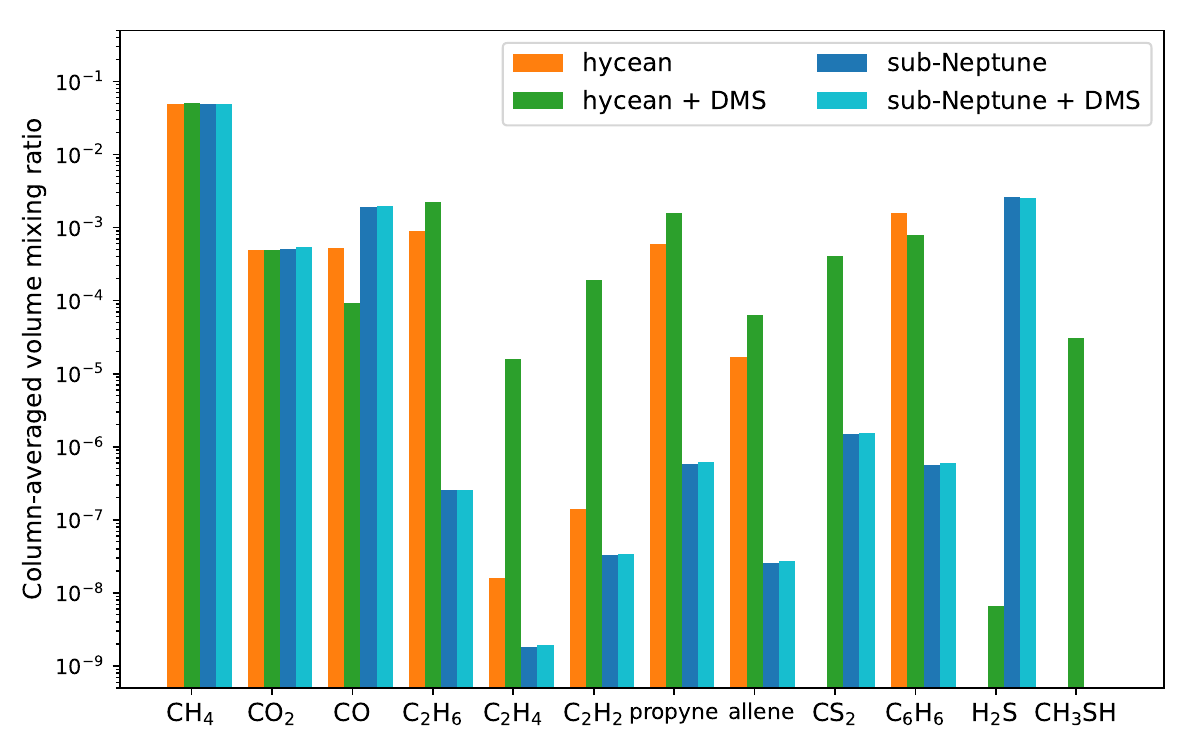}
    \caption{The column-averaged mixing ratio of key gases in the Hycean and sub-Neptune atmospheres, averaged over the pressure range of 0.01 -- 10$^{-6}$ bar to represent the expected abundances of photochemical products probed by transmission spectroscopy. \ce{CH4} and \ce{CO2} are prescribed manually at the lower boundary in the Hycean model, and the DMS distribution is also imposed for our consistency tests, as described in Sections~\ref{sec:hycean} and \ref{sec:nep}.}
    \label{fig:barcharts}
\end{figure*}

Figure~\ref{fig:barcharts} summarizes the average abundances simulated for Hycean and sub-Neptune scenarios in Sections~\ref{sec:hycean} and \ref{sec:nep}.
The key features are:
\begin{itemize}
  \item \ce{CH4} requires external sources in the Hycean scenario, whereas it is maintained by chemical equilibrium in the sub-Neptune scenario. 
  \item In a \ce{CH4}-rich environment, the conversion to higher-order hydrocarbons and aromatic compounds is highly efficient under the Hycean scenario. In this case, \ce{C2H6} and \ce{C3H4} reach $\sim$5$\%$ of \ce{CH4}, whereas they remain at $\sim$10$^{-4}$ relative to \ce{CH4} in the sub-Neptune scenario.
  \item If DMS is present in the Hycean scenario, unsaturated hydrocarbons and carbon sulfides are substantially enhanced. In particular, both \ce{C2H2} and \ce{CS2} attain abundances comparable or exceeding that of DMS. In contrast, within the sub-Neptune scenario, the presence of DMS alone does not lead to any major photochemical byproducts.
\end{itemize}

Given the robust constraints on \ce{CH4} ($\gtrsim$1$\%$), our Hycean models efficiently produce \ce{C2H6}, \ce{C3H4}, and \ce{C6H6} (as a proxy for hazes). In contrast, in the sub-Neptune scenario, \ce{C2H6} and \ce{C3H4} are produced only in the upper atmosphere and remain below $\sim$100 ppm, making them challenging to detect. The unsaturated hydrocarbons \ce{C2H2} and \ce{C2H4} reach detectable levels only in the Hycean scenario with DMS. Notably, several recent studies have pointed out that the current spectra of K2-18 b are at least equally consistent with absorption by \ce{C2H6} \citep{Luque2025} or \ce{C3H4} \citep{Welbanks2025} without invoking DMS. Additional data and analyses constraining these hydrocarbons would therefore be extremely helpful for distinguishing between the Hycean and sub-Neptune scenarios. 

While \ce{CS2} has recently been reported on several sub-Neptunes and warm giants \citep{Benneke2024,Zhang2026,Dai2026,Triantafillides2026}, it has not been identified on K2-18 b. \ce{CS2} has only a loose 2-$\sigma$ upper limit of $\sim$2\% from the combined NIRSpec and MIRI analysis of \citet{Stevenson2025}. Our simulations show that improving the upper limits of \ce{CS2} and \ce{C2H2} to $\sim$100~ppm would rule out a Hycean atmosphere with a detectable DMS abundance ($\sim$100 ppm) (Figure~\ref{fig:sen-dms}). Of the considered scenarios, the Hycean atmosphere with DMS is the only one that predicts high abundances ($\gtrsim$ 100~ppm) of \ce{C2H2} and \ce{CS2}, along with a marginally detectable level ($\gtrsim$ 1~ppm) of \ce{C2H4} and \ce{CH3SH}, and is therefore the least favored of the interpretations.

Finally, given the high abundance of \ce{C6H6} ($\sim$ 1$\%$), our model predicts that the Hycean scenario is likely to be both hazy and to retain detectable benzene, as only a small fraction of benzene ($\sim$ 50 ppm) is required to produce observable features in NIRSpec \citep{Tsai2024b}.

\subsection{Abiotic gas-phase chemistry is unlikely to produce detectable DMS}\label{sec:abiotic_dms}

\begin{figure}[t]
    \centering
    \includegraphics[width=\linewidth]{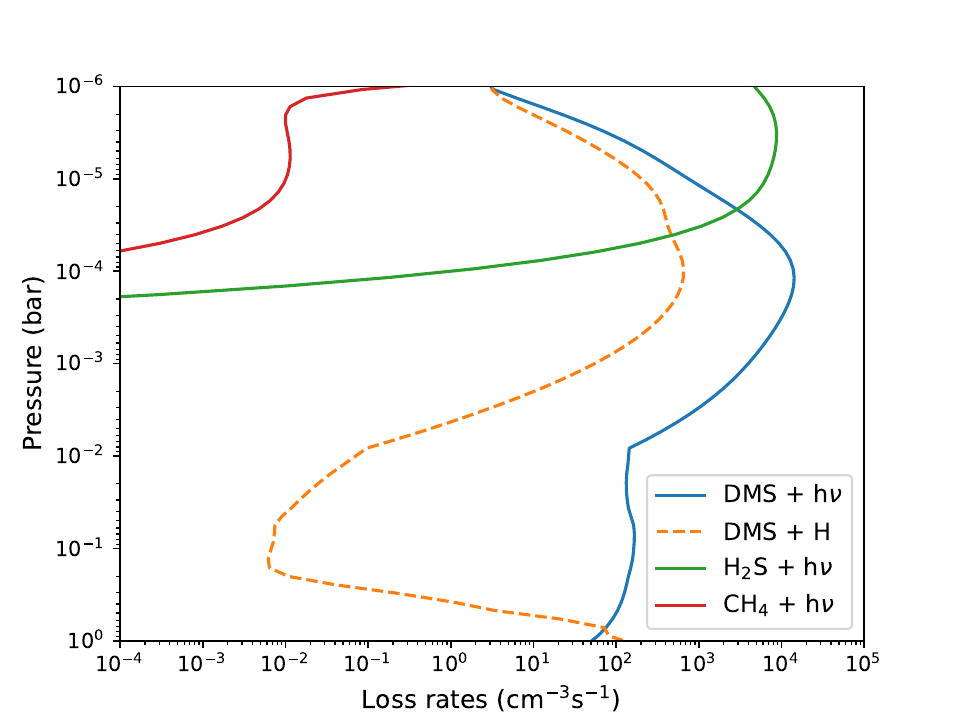}
    \caption{Photochemical loss rates of DMS in the model with a fixed Gaussian distribution (Figure~\ref{fig:nep-dms}). The photochemical loss rates of \ce{H2S} and \ce{CH4} are shown for comparison.}
    \label{fig:dms-loss}
\end{figure}


Understanding potential false positives is a central challenge in atmospheric biosignatures. Both \cite{Reed2024} and \cite{Hu2025} examine the abiotic production of DMS through purely gas-phase chemical kinetics, where the net reaction can be summarized as
\begin{equation}
\ce{2CH4 + H2S -> CH3SCH3 (DMS) + 2H2}.
\end{equation}
In both studies, DMS formation proceeds through the recombination of \ce{CH3S} and \ce{CH3} radicals. The two mechanisms differ in how \ce{CH3S} is produced. In \cite{Reed2024}, \ce{CH3S} is generated through the photolysis of \ce{CH3SH}, whereas in \cite{Hu2025}, \ce{CH3S} is built up via autocatalytic reaction cycles. A detailed comparative analysis of these pathways is presented in a parallel study \citep{Jordan2026}. In this work, we focus on a key question: 
{\it Can abiotic gas-phase chemistry, in principle, generate DMS at rates sufficient to compete against its photochemical destruction?}     

As demonstrated in Figure~\ref{fig:dms-loss}, even under the weaker UV flux of an M dwarf and with the lack of OH radicals\footnote{The OH radical is the main atmospheric sink of DMS on Earth, giving it a short lifetime of approximately one day \citep{Fung2022}. See \cite{Tsai2024b} for the comparison to \ce{H2}-atmospheres.} \citep{Tsai2024b}, photochemical destruction of DMS remains efficient in an \ce{H2}-dominated atmosphere. Beyond the gas-phase kinetics considered here, complex organic molecules are known to form through surface chemistry on grains and ices \citep{Biver2019,McKay2021}. DMS has also been identified in the coma of comet 67P/Churyumov--Gerasimenko \citep{Calmonte2016,Hanni2024}, where it may have formed through ice chemistry in the early Solar System, like other cometary organics. In all cases, regardless of its source, DMS must be produced or resupplied at a rate comparable to its photochemical sink to sustain an observable abundance.


To address this question, we explore the abiotic sulfur kinetics involving DMS production, while acknowledging the uncertainties and incompleteness in current chemical networks. As an agnostic test, we introduce a set of hypothetical reactions that provide rapid production of DMS and key intermediates -- 
\ce{CH3S} and \ce{CH3SH}:
\begin{align}
\ce{CH4 + S  &->  CH3SH}\label{eq:r1}\\
\ce{CH3 + S  &->  CH3S}\label{eq:r2}\\
\ce{CH3S + CH3 &-> CH3SCH3}\label{eq:r3}
\end{align}

\begin{figure}[t]
    \centering
    \includegraphics[width=\linewidth]{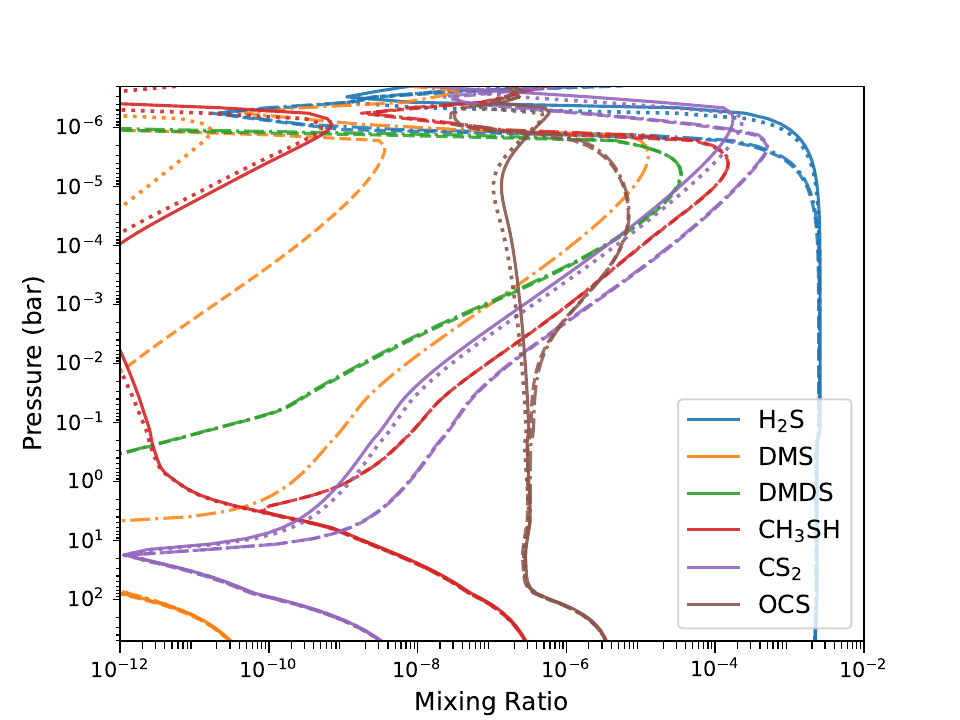}
    \caption{Important sulfur gases that are kinetically related to Earth's biogenic sulfur gases (i.e., \ce{CH3SH}, \ce{CS2}, \ce{DMS}, \ce{DMDS}, and \ce{OCS}, collectively termed S$_{\mbox{org}}$ in \citealt{Shawn2011}) in our sub-Neptune model. To test potential abiotic production of DMS in a controlled agnostic manner, the nominal abundance profile (solid) is compared with models including hypothetical reactions (\ref{eq:r1}) and (\ref{eq:r2}) (dashed), reaction (\ref{eq:r3}) (dotted), and all reactions (\ref{eq:r1})–(\ref{eq:r3}) (dash-dotted). The rate coefficients for reactions (\ref{eq:r1})–(\ref{eq:r3}) are set to the collisional upper limit, as described in Section~\ref{sec:abiotic_dms}.}
    \label{fig:dms-coll_uplim}
\end{figure}

We set the rate constants of the above reactions to the values given by the collisional frequency
\begin{equation}
k_{\mbox{coll}} = \sigma \,v_T,\quad v_T = \sqrt{\frac{8kT}{\pi \mu}}    
\end{equation}
where $\sigma$ is the impact cross section between two molecules, and $v_T$ is the relative thermal velocity, which depends on temperature and the reduced mass ($\mu$) of the colliding pair \citep{Yung1999}. This is generally an upper limit of a bimolecular reaction since it assumes all collisions result in reactions and ignores molecular interaction and electronic structures. For molecules with radii of $\sim$2~\AA, the cross section is $\sim 5 \times 10^{-15}$ cm$^2$. At 300 K, $v_T$ is $\sim 4 \times 10^{4}$ cm s$^{-1}$. This yields an upper limit for the rate coefficient of $\sim 10^{-10}$ cm$^3$ s$^{-1}$. Setting the rate coefficients of reactions (\ref{eq:r1})–(\ref{eq:r3}) to $k_{\rm coll}$ therefore serves as a limiting-case test of the conversions permitted by collisions.

Reaction~(\ref{eq:r3}) was highlighted in \cite{Reed2024} as being absent from photochemical modeling. While it is true that the forward rate of reaction~(\ref{eq:r3}) has not been experimentally quantified, the reverse reaction (i.e., the dissociation of DMS) has been measured \citep{Mousavipour2002}. In fact, reaction~(\ref{eq:r3}) was included in \cite{Tsai2024b}, with its rate derived by reversing the measured dissociation rate using thermodynamic data \citep{Visscher2011,Tsai2017}. In the deep layers of a sub-Neptune, around 1000 K and 100 bar, the resulting forward rate is approximately $10^{-12}$ cm$^{3}$ s$^{-1}$, which remains about two orders of magnitude below the collisional limit.

As illustrated in Figure~\ref{fig:dms-coll_uplim}, setting the rate of reaction (\ref{eq:r3}) to the collisional upper limit (dotted) raises DMS levels as expected, though its abundance remains negligible. When reactions (\ref{eq:r1}) and (\ref{eq:r2}) are set to the upper limit, \ce{CH3S} increases by $10^{3}$–$10^{6}$, resulting in enhanced abundances of DMS, DMDS, \ce{CH3SH}, and \ce{CS2} (dashed). DMDS shows the most significant enhancement, reaching $\sim$10 ppm through the direct recombination of \ce{CH3S}, \ce{CH3S + CH3S -> DMDS}. This confirms the role of \ce{CH3S} radical in the pathways of organic sulfur production. Nevertheless, DMS remains at a low, non-detectable abundance, with a peak mixing ratio of $\sim$1 ppt. \ce{CS2} also increases by about an order of magnitude, becoming the dominant species above $10^{-6}$ bar. Notably, we find that only when both the production of \ce{CH3S} (reactions (\ref{eq:r1})--(\ref{eq:r2})) and the \ce{CH3S}-to-DMS channel (reaction (\ref{eq:r3})) approach their collisional upper limit, does DMS reach a peak mixing ratio of $\sim$10 ppm (dash-dotted), comparable to the lower end of the abundances retrieved by \cite{Madhusudhan2023,Madhusudhan2025,PicaCiamarra2025}.


While abiotic production of DMS cannot be entirely ruled out, we argue that it is highly unlikely for reactions (\ref{eq:r1})–(\ref{eq:r3}) to operate near their collisional limits to produce potentially observable DMS above ppm-level mixing ratios. We also refer readers to \citet{Jordan2026} for discussions on the energy barriers involved in producing \ce{CH3S}.



\section{Spectral diagnostics}
\subsection{New infrared cross-section measurements of DMS and DMDS}
All of the previous studies on the DMS and DMDS detections \citep{Madhusudhan2023,Madhusudhan2025,Hu2025} rely on the cross-sections from the Pacific Northwest National Laboratories (PNNL) database \citep{PNNL2004}, which are measured with a 1-atm \ce{N2} background and available at 5, 25, and 50 $^\circ$C. Using cross sections obtained under Earth surface conditions with \ce{N2} broadening could bias the interpretation of observed transmission spectra taken in an \ce{H2}-dominated, low-pressure background, where the line width is expected to be dominated by Doppler broadening.

\begin{figure*}[htbp]
    \centering
    \includegraphics[width=0.495\linewidth]{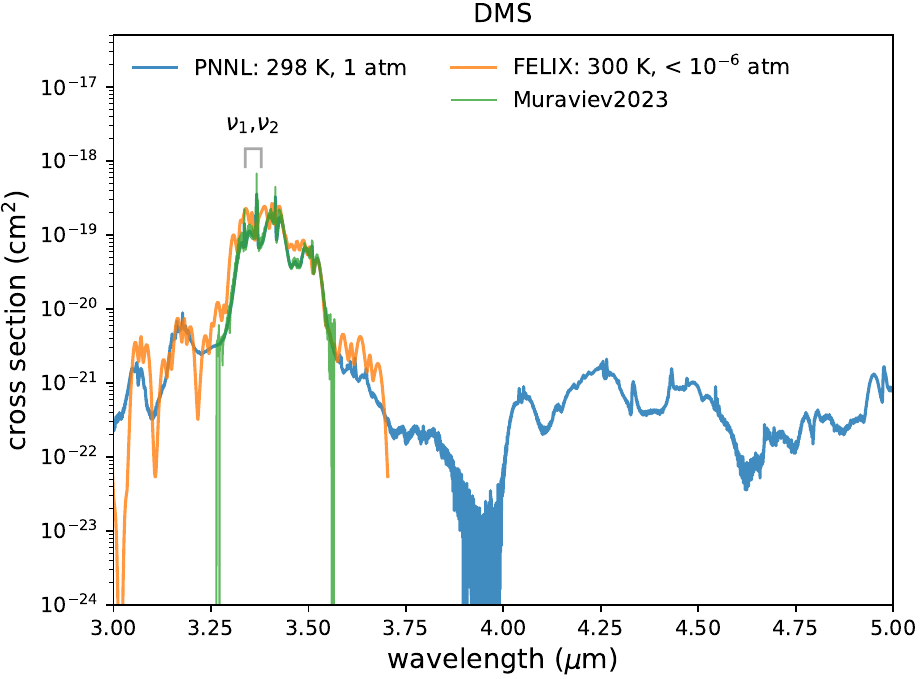}
    \includegraphics[width=0.495\linewidth]{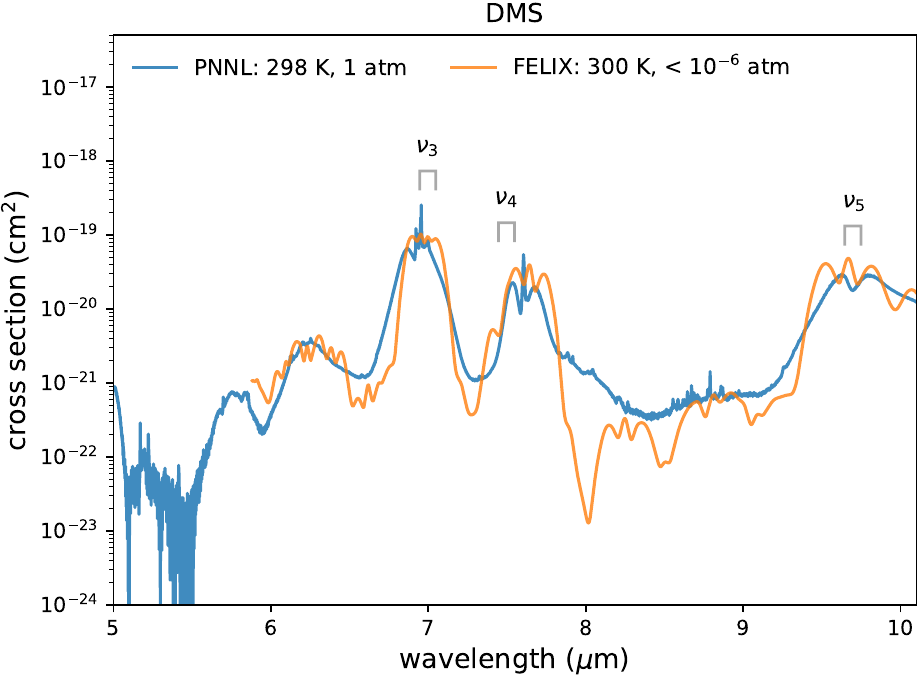}
    \includegraphics[width=0.495\linewidth]{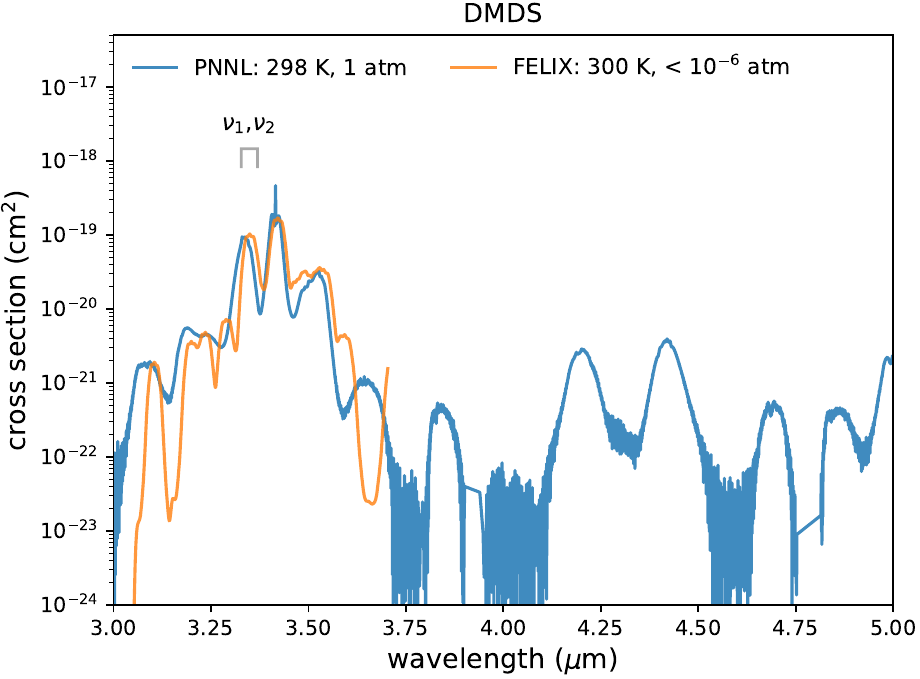}
    \includegraphics[width=0.495\linewidth]{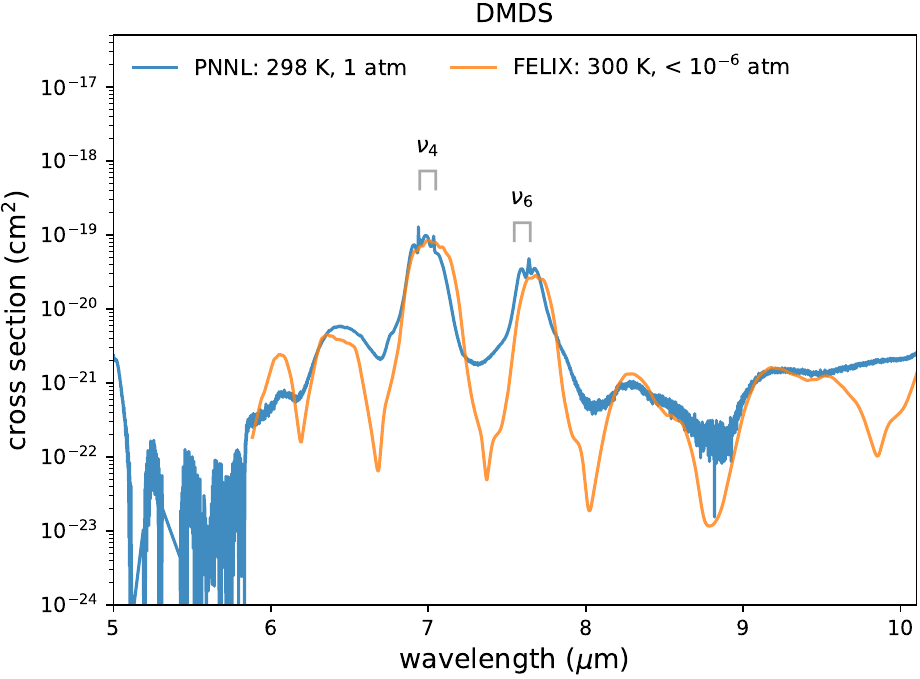}
    \caption{New semi‑experimental infrared cross sections of DMS and DMDS measured at FELIX with negligible background gas \citep{Kuipers2026}, in comparison with those from PNNL (with 1-atm \ce{N2} background) and \cite{Muraviev2023}. The strongest fundamental vibrational modes are labeled.}
    \label{fig:dms_opacities} 
\end{figure*}

Our new semi-experimental cross sections are derived from experimentally measured line positions and calculated line widths by molecular dynamics simulations, as described in Section~\ref{sec:felix}. We discard the 3.7–5.88 $\mu$m region, where no strong fundamental absorption features are robustly detected in the FELIX measurements. Figure~\ref{fig:dms_opacities} compares the resulting FELIX cross sections with those from the PNNL database. Our cross sections reproduce the main fundamental transition of both DMS and DMDS in the higher-resolution PNNL data. The narrower structures in the 6--8 $\mu$m region in the FELIX cross sections may reflect the absence of \ce{N2}-broadening or missing overtone bands due to FELIX's lower sensitivity, although the pressure-broadening effects in the PNNL cross sections for DMS and DMDS appear to be small.

\begin{figure}[htbp]
    \centering
    \includegraphics[width=\linewidth]{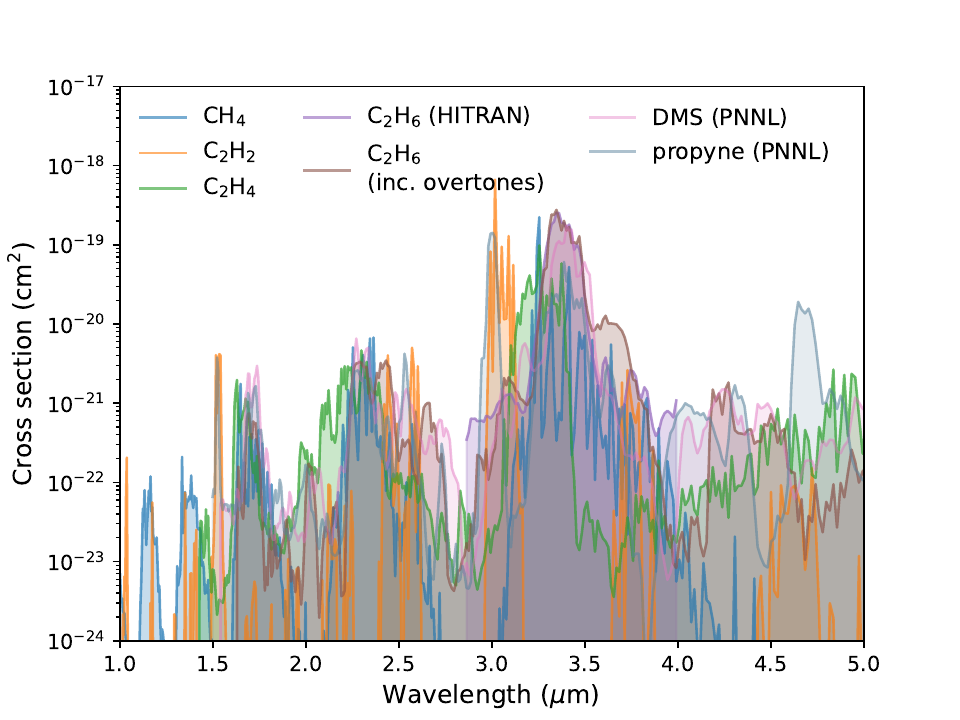}
    \includegraphics[width=\linewidth]{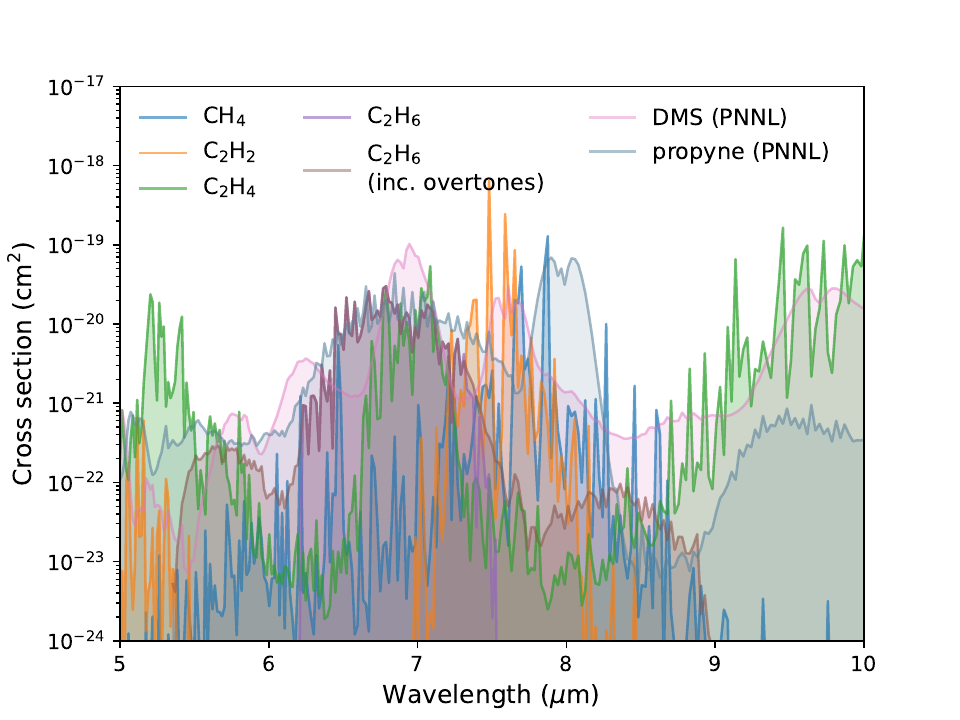}
    \caption{The comparison of major hydrocarbon and DMS opacities in the near-IR (top) and mid-IR (bottom) ranges. Cross sections are shown at 300 K and 1 mbar, except for the PNNL data, which were measured in 1-atm \ce{N2}.} 
    \label{fig:opacities}
\end{figure}

\begin{figure}[htbp]
    \centering
    \includegraphics[width=\linewidth]{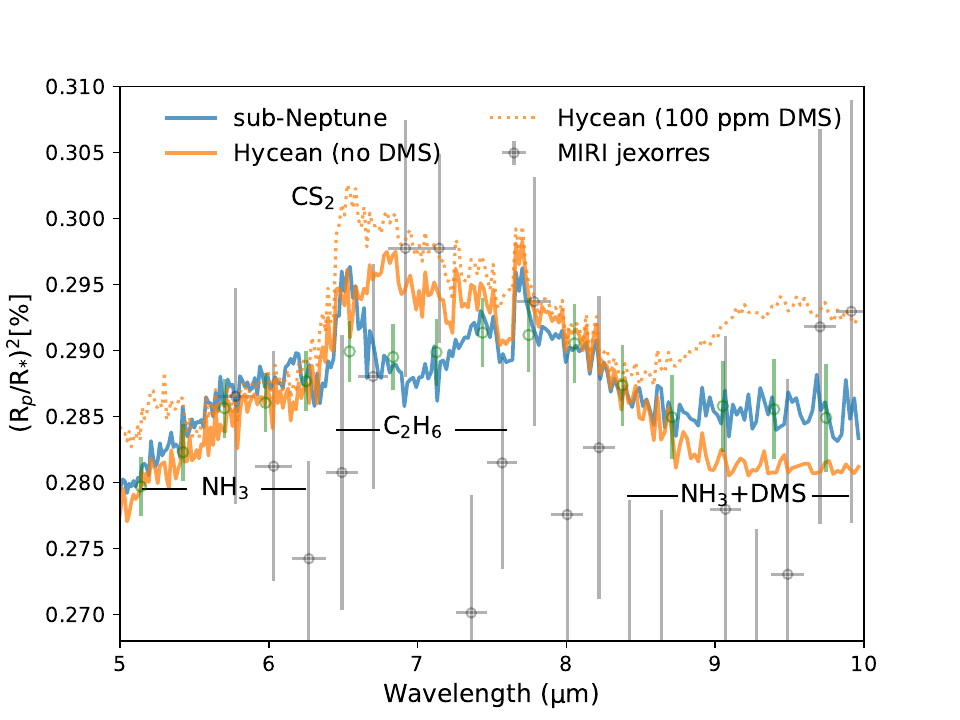}
    \caption{Synthetic transmission spectra of our sub-Neptune versus Hycean models. Green error bars show simulated noise by Pandexo with the JWST/MIRI LRS for 3 transits.}
    \label{fig:spec_nep_hycean}
\end{figure}

\subsection{Constraining simple hydrocarbons with different wavelength windows}
Recent analyses have shown that in JWST transmission spectra of fainter targets, such as sub-Neptunes around M dwarfs, molecular features can be degenerate among multiple species \citep[e.g., ][]{Gressier2024,Banerjee2024,Welbanks2025,Luque2025,Stevenson2025}. The degeneracy is particularly severe with methane and hydrocarbons that share the same C-H fundamental band (e.g., see Figure 3 in \cite{Niraula2025}). We now discuss the optimal observational window to constrain simple hydrocarbons to better interpret the planetary context examined in Section~\ref{sec:dms}. 

We compare the opacities of stable (closed-shell) hydrocarbons, \ce{C2H2}, \ce{C2H4}, \ce{C2H6}, \ce{CH3CCH} (propyne), with \ce{CH4} and DMS overlaid in Figure~\ref{fig:opacities}. The cross sections data are adopted from \cite{Rey2017} for \ce{CH4}, HITRAN \citep{Gordon2022} for \ce{C2H2}, and ExoMol \citep{Tennyson2024} for \ce{C2H4}. For \ce{C2H6}, the near-infrared line lists remain incomplete for planetary applications \citep[e.g.,][]{Schwieterman2018}. Early line lists of \ce{C2H6} in HITRAN included only the $\nu_7$ stretching band near 3.3 $\mu$m and the $\nu_8$ deformation band near 6.8 $\mu$m. In HITRAN2020 \citep{Gordon2022}, the \ce{C2H6} line list was updated based on \citep{Lattanzi2011,Villanueva2011} with air-broadening. In the recent HITRAN2024 release \citep{Gordon2026}, both \ce{C2H2} and \ce{C2H6} have been further updated, including the 3.8 $\mu$m region for \ce{C2H2} and corrections to the air- and self-broadening parameters of \ce{C2H6}. To better represent \ce{H2}-dominated exoplanet atmospheres, we compile the experimental cross sections of \ce{C2H6} from \cite{Hewett2020} and \cite{Dodangodage2020}, which provide high-resolution measurements with \ce{H2}-broadening. \cite{Hewett2020} aimed to resolve weak overtone and combination bands between 1.6 and 5.5 $\mu$m, whereas \cite{Dodangodage2020} measured the strong C–H fundamental bands near 3.3 $\mu$m, which were saturated in the measurements of \cite{Hewett2020}. In addition, the PNNL cross sections include the overtone and combination bands across the 5--9 $\mu$m region. We therefore adopt the absorption cross sections from \cite{Hewett2020} for $\lambda < 3.24$ $\mu$m, from \cite{Dodangodage2020} for $3.24 \le \lambda \le 5.4$ $\mu$m, from PNNL for $5 \le \lambda \le 6.2$ $\mu$m and $7.3 \le \lambda \le 9$ $\mu$m, and from HITRAN2020 for $6.2 \le \lambda \le 7.3$ $\mu$m. For propyne, as for DMS, there is currently no available line list covering the relevant wavelength range, and we adopt the PNNL cross sections measured at 298~K in 1~atm of \ce{N2}. These likely overestimate pressure broadening at the mbar-level pressure.
        
We find that \ce{C2H2} is best characterized by its fundamental C-H stretching band near 3 $\mu$m, where it is relatively free from overlap with the other C$_2$ hydrocarbons.  Its 3-$\mu$m feature overlaps with propyne but could be distinguished by propyne's C$\equiv$C stretching band at 4.7 $\mu$m. \ce{C2H4} may be constrained in the 5–5.5 $\mu$m range, but it only reaches detectable abundances in the Hycean scenario with DMS. \ce{C2H6} is a saturated hydrocarbon and generally the most abundant C2-hydrocarbon on cool sub-Neptunes. Our compiled experimental \ce{C2H6} cross sections agree well with the line lists from HITRAN2020 at the strong 3.3 $\mu$m fundamental band, whereas the weak overtones in \cite{Hewett2020} are not included in HITRAN2020. These overtone bands appear to overlap with DMS features, while the strong 3.3-$\mu$m mode overlaps with both DMS and \ce{CH4}. Hence, the most favorable window for distinguishing \ce{C2H6} is likely its bending band around 6.8 $\mu$m. Propyne's features at 4.7~$\mu$m and near 8~$\mu$m are relatively free from the other hydrocarbons and DMS, but the former overlaps with CO and the latter partly overlaps with \ce{CH4}. With only a single transit, \citet{Welbanks2025} found that the current MIRI spectrum cannot distinguish propyne from DMS in the 6--8~$\mu$m range.


Finally, we compare the synthetic transmission spectra produced by our nominal sub-Neptune and Hycean (with and without DMS) models in Figure~\ref{fig:spec_nep_hycean}. To recap, \ce{CH4} and \ce{CO2} are produced self-consistently under chemical equilibrium in the sub-Neptune scenario, whereas in our controlled Hycean model, they are prescribed at the lower boundary. There are three main diagnostic features for distinguishing the two scenarios. First, the sub-Neptune atmosphere is expected to retain \ce{NH3}, whereas \ce{NH3} is efficiently destroyed in the Hycean case \citep{Yu2021,Tsai2021b,Hu2021}. While the detection of ammonia would be a robust indicator of the sub-Neptune scenario, its non-detection would not uniquely favor a Hycean interpretation, as nitrogen may dissolve into a reduced deep magma reservoir \citep{Shorttle2024,Glein2025}. Second, the Hycean scenario predicts a notable \ce{C2H6} feature near 6.8 $\mu$m in the presence of \ce{CH4}. Third, if DMS is present in the Hycean scenario at the reported abundance ($\sim$100~ppm), its photochemical byproduct \ce{CS2} produces a notable feature near 6.6~$\mu$m. Both the feature differences between the Hycean and sub-Neptune scenarios and between the Hycean scenario with DMS and the sub-Neptune scenario are approximately 50 ppm in transit depth, which is below the sensitivity of a single MIRI/LRS transit but might become detectable with multiple transits, as illustrated in Figure~\ref{fig:spec_nep_hycean}.






\begin{figure}[t]
    \centering
    \includegraphics[width=\linewidth]{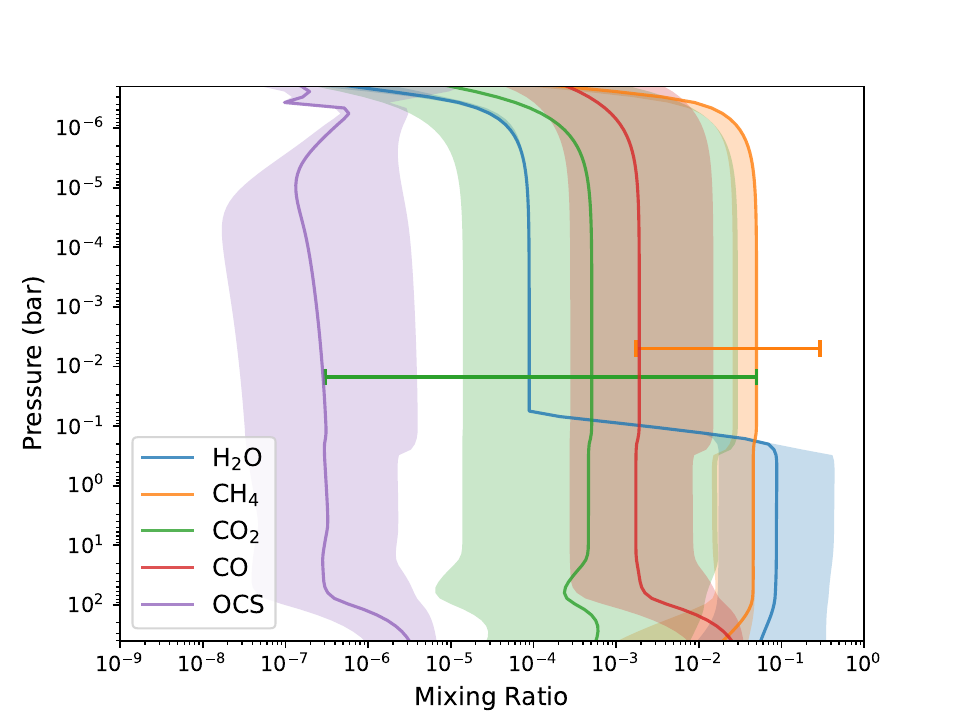}
    \caption{The shaded regions present the range of mixing ratios with the oxygen enhancement factor varying from 0.2 to 5 from the nominal sub-Neptune model. The nominal case (100$\times$ solar metallicity) is shown as the solid lines. Same error bars as in Figure~\ref{fig:apo} show the combined 1-$\sigma$ ranges from multiple retrieval analyses listed in Table~\ref{tab:retrievals}.  
    }
    \label{fig:O5X}
\end{figure}
    
\section{Discussion}\label{sec:discussions}
\subsection{K2-18 b does not require an H$_2$O-rich interior}
Atmospheric characterization potentially offers a window to distinguish between the gas-dwarf scenario and the water-world scenario for sub-Neptunes. \cite{Yang2024} demonstrated that the atmospheric \ce{CO2}/\ce{CH4} ratio serves as an informative proxy for tracing the bulk \ce{H2O} content, with \ce{SO2} and OCS offering additional constraints. We apply this framework to K2-18 b by varying the oxygen abundance in our nominal 100$\times$ solar metallicity model by factors of 0.2 and 5, which we refer to as the oxygen enhancement factor. The 0.2 and 5 oxygen enhancement corresponds to approximately the deep \ce{H2O}/\ce{H2} of 2:100 and 50:50 used in \cite{Yang2024,Hu2025}. Given the current observational uncertainties (Table~\ref{tab:retrievals}), models with this range of oxygen enhancement are all consistent with the existing data, as shown in Figure~\ref{fig:O5X}. To meaningfully constrain the bulk oxygen content, the abundance of \ce{CO2} would need to be determined to within approximately an order of magnitude. As our nominal sub-Neptune model with 100$\times$ solar metallicity is consistent with observational constraints, except for the non-detection of \ce{NH3}, we find no need to invoke an enhanced \ce{H2O}-rich interior to explain K2-18 b.




\subsection{Constraining eccentricity}
Current constraints on the orbital eccentricity of K2-18 b are limited. The analysis by \cite{Sarkis2018} relies on the small modulation of stellar radial velocity, on a similar order of magnitude to the stellar activity jitter. In addition to obtaining more radial velocity, we briefly discuss other independent methods to help constrain the orbital eccentricity of temperate sub-Neptunes.

One way to constrain orbital eccentricity is to measure the timing and duration of the secondary eclipse relative to the transit \citep{Alonso2018}. Each of these measurements is degenerate with the argument of periastron, so combining both measurements can help break this degeneracy and better constrain the eccentricity. The upside is that the long orbital periods of temperate sub-Neptunes amplify the timing offset. The downside is that the secondary eclipses of temperate sub-Neptunes such as K2-18~b are expected to be very shallow. Measuring their timing and duration precisely is therefore challenging, especially because the uncertain eccentricity also makes it uncertain when the eclipse would occur, producing an ambiguity as to whether a non-detection is due to a shallow eclipse or a missed eclipse.


If there is a second planet in this system, such as K2-18~c \citep{Cloutier2017}, whose RV interpretation has been disputed \citep{Sarkis2018} but later supported by reanalyses \citep{Cloutier2019,Radica2022}, it may cause transit duration variation (TDV) and/or transit timing variation (TTV) of the transiting planet. If these are detected and measured, they could be used to constrain the eccentricity. However, the constraints would be degenerate with the mass determination of the star and the planets unless these can be obtained via other methods. Additionally, if the perturbing companion is non-transiting, there are likely to be multiple viable solutions for the eccentricity. 

Alternatively, one could constrain the eccentricity using high-resolution cross-correlation spectroscopy (HRCCS) technique \citep{Birkby2018,Snellen2025}. This technique requires a several-hour time series of high-signal-to-noise stellar spectra to extract the planet's faint, high-resolution spectrum. The observations also provide a finely sampled time series of stellar radial velocities. Because these measurements span a short time interval, they may be less susceptible to stellar activity and could be used to constrain the eccentricity via the radial-velocity method. Additionally, the detection of the planet's spectrum with HRCCS would yield a measurement of the velocity of K2-18 b. Since winds on K2-18 b are expected to be slow ($\lesssim100$ms$^{-1}$ \cite[e.g., ][]{Tsai2024b}), any offset in the measured velocity could reflect errors in the assumed orbital parameters and thus help constrain the eccentricity. As with secondary-eclipse timing, this constraint would remain degenerate with the argument of periastron. Nevertheless, it could improve predictions of the secondary-eclipse time and thereby aid its detection, allowing the eccentricity to be further constrained.

Taken together, further constraining the eccentricity of K2-18 b may require a combination of techniques. We suggest first using HRCCS to improve the radial-velocity constraints on the eccentricity and refine the predicted secondary-eclipse time, followed by observations to detect and measure the timing and duration of the eclipse.

\subsection{Stability of orbital eccentricity}
According to \cite{Gomes2020}, the secular evolution of the K2-18 b-c system is governed by the physical properties of the inner planet, K2-18 c. Specifically, the key parameter is the planet's relaxation factor -- how quickly a tidally deformed body adjusts to its equilibrium shape under tidal perturbation (i.e. zero relaxation factor corresponds to no deformation for a perfectly rigid solid body). If K2-18 c has a rocky interior and a small relaxation factor, the orbits of both planets are expected to be circularized within $\sim$0.1 Gyr. If K2-18 c is a gaseous mini-Neptune with a large relaxation factor, moderately high eccentricity (e.g. about 0.2) can be sustained for $\sim$ Gyrs, consistent with the age of the system.





\subsection{Does water condensation driven by orbital eccentricity persist throughout the orbit?}
Since we did not explicitly model the temporal evolution of the climate, one question that arises is whether the condensation of water near apoastron can account for observations obtained at different orbital phases. As shown in Figure~\ref{fig:bifur}, under the stellar irradiation within the bifurcation region, \ce{H2O} clouds are expected to persist moving out from apoastron but do not form from periastron. 

We argue, however, that after the clouds dissipate, the recovery of water above the cloud top requires vertical mixing to transport water from the deep layers. This process operates on the timescale of atmospheric dynamics, even when clouds dissipate over a short timescale. In Earth’s upper troposphere and lower stratosphere, the mixing timescale is on the order of a few months \citep{Minschwaner1996} (effective $K_{zz}$ = 10$^4$--10$^5$ cm$^2$s$^{-1}$). For a temperate sub-Neptune, a moderate eddy diffusion of $K_{zz}$ = 10$^7$ cm$^2$s$^{-1}$ used in our K2-18 b model, also leads to a mixing timescale of about a month, comparable to the orbital period of K2-18 b, as well as typical orbital periods of temperate sub-Neptunes around M dwarfs. Although vertical mixing of K2-18~b is not directly constrained and instead informed by circulation models, water cold trapping could remain effective throughout most of the orbit, provided that vertical mixing does not exceed the eddy diffusion coefficient, \(K_{zz}=10^7\) cm\(^2\) s\(^{-1}\), adopted in our nominal model. Future surveys to probe atmospheric water abundance on eccentric sub-Neptunes with average instellation comparable to Earth can test the robustness of the water cold trapping mechanism.

\begin{table*}[!t]
\centering
\begin{tabular}{c|c|c|l}
 & Hycean & Sub-Neptune & \multicolumn{1}{c}{Notes}\\
\hline
\ce{CH4} & $<1$ ppm\footnote{\cite{Wogan2024,Tsai2024b}} & $10^{-2}$--$10^{-1}$ & \\
\ce{CO2} & $4\times10^{-4}$--$10^{-1}$\footnote{\cite{Hu2021}} & $10^{-5}$--$10^{-2}$ &\\
\ce{C2H6} & $>$1000 ppm & $<$1000 ppm & spectral window: 6–8 $\mu$m\\
\ce{C2H2} & $\sim$100 ppm with DMS& $<$100 ppm & spectral window: 3 $\mu$m\\
\ce{C2H4} & $\sim$100 ppm with DMS & $<$100 ppm & spectral window: 5–5.5 $\mu$m\\
\ce{NH3} & negligible & $\sim$1000 ppm & sensitive to magma/water oceans\\
\ce{CS2} & $>$100 ppm with DMS & 1--100 ppm & spectral window: 6.6–6.7 $\mu$m\\
\end{tabular}
\caption{Summary of atmospheric abundances predicted by photochemical modeling in this study and the literature for both Hycean and sub-Neptune scenarios.}
\label{tab:summary}
\end{table*}

\section{Conclusions}\label{sec:conclusions}
This study addresses three outstanding physical questions about K2-18 b by applying a coupled climate–photochemistry framework to reassess its atmospheric composition. We adopt an agnostic approach to test whether the inferred abundances from spectral retrievals are self-consistent for both Hycean and sub-Neptune interpretations. First, we show that water clouds induced by a mildly eccentric orbit naturally strengthen the water cold trap and can physically account for the observed low water abundance without invoking ad hoc reductions in stellar flux. Second, our photochemical modeling demonstrates that in a \ce{CH4}-rich Hycean atmosphere, the conversion to higher-order hydrocarbons and aromatic precursors is highly efficient. Because a shallow Hycean atmosphere lacks a deep thermochemical reservoir, the observed \ce{CH4} can only be maintained by a sustained surface flux \citep{Wogan2024}, and its photochemical processing inevitably leads to the accumulation of higher-order hydrocarbons. Further observations to detect or rule out hydrocarbons, particularly \ce{C2H6} and \ce{C3H4} (propyne), can help confidently discriminate between Hycean and sub-Neptune scenarios. Third, we find that abiotic gas-phase chemistry is unlikely to produce DMS at observable levels: only when the relevant reactions are pushed to their collisional limits does DMS exceed ppm-level abundances at high altitudes. We summarize the key molecular diagnostics identified in this work and the literature in Table~\ref{tab:summary}. Overall, our results favor a high-metallicity sub-Neptune atmosphere for K2-18 b without invoking DMS, and we find that current observational constraints permit a wide range of interior water contents.

\begin{acknowledgments}
The authors thank Robert Hargreaves and Pei-Ling Luo for advice on the \ce{C2H6} and DMS opacities; Romeo Veillet for discussions on the DMS formation pathways; Julian Moses for discussions on hydrocarbon UV cross sections and the chemical pathways of \ce{CS2}; Michael Zhang for inspiring suggestions on model comparisons; Kevin Stevenson for helpful discussions and for sharing the retrieval results; and Daria Galimberti for support with quantum chemical calculations. 
S.-M. T. is supported by the National Science and Technology Council (grants 114-2112-M-001-065-MY3) and an Academia Sinica Career Development Award (AS-CDA-115-M03). L.P. and B.O. acknowledge funding by the Helmholtz Initiative and Networking Fund. B.O. acknowledges support from the BMBF (Grant No. 13 K22CHA). S.J. acknowledges funding support from ETH Zurich and the NOMIS Foundation in the form of a research fellowship. J.L.Y. and E.W.S. acknowledge funding support from the CHAMPs (Consortium on Habitability and Atmospheres of M-dwarf Planets) team, issued by the National Aeronautics and Space Administration (NASA) under Grant No. 80NSSC23K1399 through the Interdisciplinary Consortia for Astrobiology Research (ICAR) program. 
The authors gratefully acknowledge the Nederlandse Organisatie voor Wetenschappelijk Onderzoek (NWO) for the support of the HFML-FELIX Institute and for CPU time on the Dutch National Supercomputer Snellius (Project 2024.009), as well as the constant support of the HFML-FELIX staff. 
AI tools (Claude Opus 5.5) were used to assist with proofreading the final manuscript.


\end{acknowledgments}

\bibliography{master_bib}
\bibliographystyle{aasjournalv7}



\end{document}